\documentclass[lettersize,journal]{IEEEtran} 
\IEEEoverridecommandlockouts
\makeatletter
\def\ps@headings{%
\def\@oddhead{\mbox{}\scriptsize\rightmark \hfil \thepage}%
\def\@evenhead{\scriptsize\thepage \hfil \leftmark\mbox{}}%
\def\@oddfoot{}%
\def\@evenfoot{}}
\makeatother
\usepackage{amsthm}
\makeatletter
\def\th@plain{%
\thm@notefont{}
\itshape 
}
\def\th@definition{%
\thm@notefont{}
\normalfont 
}
\makeatother
\usepackage{dsfont}
\usepackage[table]{xcolor}  
\usepackage{graphicx}
\usepackage{subcaption}

\usepackage{amsfonts}
\usepackage{cite}
\usepackage{epstopdf}
\usepackage{mathrsfs}
\usepackage{graphicx}
\usepackage{amsmath}
\usepackage{times}
\usepackage{bm}
\usepackage{multirow}
\usepackage{latexsym}
\usepackage{amssymb}
\usepackage{caption}
\usepackage{stfloats}
\usepackage{cases}
\usepackage{amssymb}
\usepackage{xcolor}
\usepackage{array}
\usepackage{setspace}
\usepackage{fancyhdr}
\usepackage{graphicx}
\usepackage{balance}
\usepackage{threeparttable}%
\usepackage{mathtools}
\usepackage{enumerate}
\usepackage[T1]{fontenc}
\usepackage{aecompl}
\usepackage{booktabs}
\usepackage{graphicx} %
\usepackage{cuted}
\usepackage[ruled,linesnumbered]{algorithm2e}
\usepackage{algpseudocode}
\usepackage[mathstyleoff]{breqn}
\usepackage{makecell}
\usepackage{verbatim}%

\DeclareCaptionLabelFormat{modified}{\text{Fig.\ #2}}
\begin{document}
\captionsetup[figure]{name={Fig.},labelsep=period,singlelinecheck=off}
\title{Joint Age and Coverage-Optimal Satellite Constellation Relaying in Cislunar Communications with Hybrid Orbits
\\
\thanks{A. Yuan, Z. Hu, Q. Zhang, Z. Yang* are with the Communication Engineering Research Center, Harbin Institute of Technology, Shenzhen, 518055, China. (e-mail: yuanaf@stu.hit.edu.cn, huzhouyong@stu.hit.edu.cn, zqy@hit.edu.cn, yangzhihua@hit.edu.cn)}
\thanks{Z. Sun is with the Institute of Communication Systems, University of Surrey, GU2 7XH Guildford, U.K. (e-mail: z.sun@surrey.ac.uk)}
\thanks{Z. Yang* and Q. Zhang are with the Pengcheng Lab, Shenzhen, 518055, China. (Corresponding author: Zhihua Yang, email: yangzhihua@hit.edu.cn)}
\thanks{This work is supported in part by the National Natural Science Foundation of China under Grant 62371158, the Major Key Project of PCL(PCL2024A01), and China Scholarship Council.}} 
\author{\IEEEauthorblockN{~Afang Yuan, ~Zhouyong Hu, {~Zhili Sun, \textit{Senior Member, IEEE}}, \\{~Qinyu Zhang, \textit{Senior Member, IEEE}}, and~Zhihua Yang* \\}}

\maketitle
\IEEEpeerreviewmaketitle

\begin{abstract}
With the ever-increasing lunar missions, a growing interest develops in designing data relay satellite constellations for cislunar communications, which is challenged by the constrained visibility and huge distance between the earth and moon in pursuit of establishing real-time communication links. In this work, therefore, we propose an age and coverage optimal relay satellite constellation for cislunar communication by considering the self-rotation of the earth as well as the orbital motion of the moon, which consists of hybrid Earth-Moon Libration 1/2 (EML1/L2) points Halo orbits, ordinary lunar orbits, and Geostationary Earth Orbit (GEO) satellites. In particular, by minimizing both the number of satellites and the average per-device Age of Information (AoI) while maximizing the coverage ratio of specific lunar surface regions, a multi-objective optimization problem is formulated and solved by using a well-designed Nondominated Sorting Genetic Algorithm-II (NSGA-II). The simulation results demonstrate that our proposed hybrid constellation significantly outperforms traditional Walker Star and Delta constellations in terms of both AoI and the coverage of communication.
\end{abstract}

\vspace{0.15cm}
\vspace{0.1cm}

\IEEEpeerreviewmaketitle

\section{Introduction}
\subsection{Motivation}
Lunar exploration has become one of current hot spots in international space exploration, with major space agencies initiating lunar missions. The United States National Aeronautics and Space Administration (NASA) proposed the Artemis mission to return humans to the moon and build long-term infrastructure in orbit and on the surface in 2017 \cite{a1, a2, a3}. The European Space Agency (ESA) launched the Moonlight initiative in 2020 to stimulate the creation and development of lunar communication and navigation services \cite{a6}. China has completed the first three phases of the lunar exploration project for orbiting, landing and returning missions \cite{a7,a8}, and is carrying out the fourth phase of the lunar exploration project to construct an international lunar scientific research station \cite{a9,a10,a11}. At the same time, Russia, Japan, and India are also actively carrying out lunar exploration missions \cite{a12,a13}.
Lunar missions are increasingly complex, from unmanned exploration and scientific research station to manned lunar landings, and need efficient and real-time communication system as the infrastructure to support these complex missions and enhance reliability. However, the significant distance from earth to the moon, approximately 380,000 km away, prevents lunar probes with heavily limited power from sending information directly back to earth, necessitating intermediary communication systems. In addition, the Lunar Far Side (LFS) continuously faces away from the earth due to the moon's synchronous rotation, hindering direct communication links between the LFS and earth. Given the anticipated diversity of lunar missions, a combined relay satellite network supporting real-time communication between earth and moon is necessary to ensure comprehensive communication capabilities.

Currently, there are interesting works on designing Earth-Moon relay satellite constellations, which have following limitations requiring to be addressed. Firstly, ordinary lunar orbital satellites are focused on and exploited as design objective, i.e., lunar elliptical orbit or circular orbit satellite, which do not take full advantage of the Earth-Moon Libration Point (EMLP) orbit that has attracted broad interest in lunar exploration missions such as berthing transit, astronomical observation, and relay communication. Achieving full lunar-surface coverage requires numerous ordinary lunar relay satellites, which may not be practical to meet the requirements for the lunar exploration mission. Besides, traditional latency or propagation delay has been considered as an important metric for evaluation of data transmission quality of service (QoS). Nowadays, as an information freshness metric, Age of Information (AoI) captures both the latency and the generation time of each status update from the receiver's perspective, making it better suited and effective than the traditional delay metrics for representing the timeliness of Earth-Moon space systems. Additionally, the coverage performance of the lunar surface is overemphasized by optimization methods, neglecting the actual motions of the moon’s orbit and rotation. In realistic, considering the complexity of the Earth-Moon motion system, simplifying the earth and moon as two static celestial bodies will negatively impact the effectiveness of satellite constellation, leading to the incapability of real-time communications between the earth control center of lunar mission and the deployed constellations covering the entire lunar surface.

Therefore, considering the economic factors of satellite deployment and the complex relative motion between the earth and moon, integrating EMLP orbits with ordinary lunar orbits is an urgent issue that needs to be addressed to achieve real-time Earth-Moon communication constellations.

\subsection{Contributions}
Despite a few works making efforts on the design of satellite constellations targeting the entire lunar surface, there are scarce results considering the relay constellation design with hybrid orbits from the perspective of information timeliness and economic costs. In this paper, we design a combined Geostationary Earth Orbit (GEO), ordinary lunar orbit, and EMLP Halo orbit satellites constellation system by jointly optimizing the number of satellites, information freshness and coverage for the lunar exploration missions considering the rotation of the earth and the revolution of the moon. The main contributions of this paper are summarized as follows:
\begin{itemize}
\item We develop an age and coverage optimal integrated Earth-Moon relay satellite constellation for cislunar communication, which consists of hybrid Earth-Moon Libration 1/2 (EML1/L2) points Halo orbits, ordinary lunar orbits, and GEO satellites. In particular, a unified coordinate system is analytically employed to achieve the complementary coverage of different orbital types by transforming the lunar coordinates of Earth-Moon system into the Earth-Centered Inertial (ECI) frames.

\item We formulate the optimization problem of constellation design as a multi-objective optimization problem, aiming to minimize the total number of satellites as well as the average per-device AoI and to maximize the coverage of a specified lunar region, which is solved by using a well-designed Nondominated Sorting Genetic Algorithm-II (NSGA-II) as well as a grid-point analysis method to determine the optimal configuration.

\item The simulation results show that the AoI and coverage performance of our proposed constellation design is significantly better than that of the traditional Walker Star and Delta constellations. According to our limited knowledge, this is the first work to investigate the hybrid constellation design of cislunar communication with respect to the timeliness, which will be beneficial to the development of space communications techniques and lunar explorations.
\end{itemize}

The rest of this paper is structured as follows. Section II introduces the related work. In Section III, preliminary fundamentals are presented. The problem is formulated with the system model in Section IV. In Section V, multi-objective optimization algorithm is specified. Section VI provides the performance analysis. Finally, Section VII gives the conclusion and directions for future research.

\vspace{-0.05cm}
\section{Related Work}
In recent years, most of the works \cite{a18, a20,a23, a24, a26,a25,a27,a28,a30,a17,a21,25a,a29,a31,a32,a33} have investigated lunar relay communication networks from the perspective of coverage. The authors in \cite{a18} investigate multiple-orbit constellation communication networks and state the trajectory design methods in single and multiple polar orbit planes. In \cite{a20}, an enhanced orbital design is presented to ensure superior south pole coverage by applying lunar frozen orbits. The theoretical minimum number of satellites required for various lunar coverage scenarios at a specific altitude is proposed in \cite{a23}, which also delineates the design criteria for communication and navigation. The study in \cite{a24} introduces two tilted elliptical constellations with six satellites to achieve comprehensive lunar polar and global coverage. The authors in \cite{a26} design a small lunar navigation and communication satellite system with earth-global positioning system time transfer that provides services near the lunar south pole. Additionally, two constellations providing 99.999\% global coverage without orbit maintenance are designed in  \cite{a25}. In \cite{a27}, the authors investigate Pareto-optimal lunar communication-navigation integrated satellite constellations with low latency in lunar frozen orbit space based on the difference analysis of earth-moon constellation construction. The work in \cite{a28} investigates lunar satellites to identify the optimal parameters to broadcast to a lunar user.  The authors in \cite{a30} investigate the Lunar Pathfinder data relay satellite and its orbit - an Elliptical Lunar Frozen Orbit (ELFO). The authors in \cite{a17} analyze the coverage ability of Halo orbits and Distant Retrograde Orbit (DRO) and investigate constellation combinations with different numbers of satellites. The study in \cite{a21} develops a coverage evaluation algorithm to assess both instantaneous and global lunar coverage, designing a constellation of three satellites in Halo orbits near the EML1/L2 points. The study in \cite{25a} analyzes multiple types of orbits and proposes a hierarchical communication architecture and network protocol for the cislunar constellation. The authors in \cite{a29} present a highly evolvable, low-cost lunar relay constellation concept using small satellites in EML1/L2 Halo orbits. In \cite{a31}, the authors analyze different types of orbits and propose an orbit design method for near-moon space constellation for lunar communication and navigation. In \cite{a32}, the authors realize a lunar global positioning system using north and south families of Halo orbits around  EML1/L2 libration points. The work in \cite{a33} presents four different types of constellations consisting of the multi-revolution elliptic Halo orbit. However, these studies do not consider the design of heterogeneous constellations and the combined Earth-Moon motion, and only consider constellation design from the lunar side.

To better quantize timeless of information, the AoI-oriented optimization of different networks has been widely investigated in \cite{b26,b27,b28, b29,b30,b31,a35, a37, a38, a39, a40}. The research in \cite{b26,b27,b28, b29,b30,b31} consider multi-source single-hop networks and \cite{ a35,a37, a38, a39, a40} take into account AoI optimization in multi-hop networks. However, their work does not consider the sources' coverage situation of the relay. To the best of the authors' comprehension, no existing works focus on the satellite constellation design using EML1/L2 Halo orbits, ordinary orbits and earth orbit to improve the information freshness and coverage ratio considering the rotation of the earth and the revolution of the moon.

\section{Preliminary Fundamentals}
\subsection{Earth-Moon Coordinate System}

To evaluate the performance of the time-varying Earth-Moon system caused by the earth's rotation and the movement of the moon and satellites, we consider different Earth-Moon coordinate system transformations and use the J2000 ECI coordinate frames as the unified coordinate system. The detailed coordinate transformation process will be introduced below: Earth-centered Earth-fixed (ECEF) Transform to ECI and Lunar Center Equatorial (LCE) Transform to ECI.

\subsubsection{ECEF Transform to ECI}

The ECEF coordinate frame is commonly used to describe the location of an object on earth. ECEF coordinate frame rotates with the earth, with its $x$-$y$ plane corresponding to the earth's equatorial plane. The $x$-axis intersects the earth's sphere at $0^\circ$ latitude (equator) and $0^\circ$ longitude (Greenwich meridian), the $y$-axis points towards $90^\circ$ east longitude, and the $z$-axis points towards the north pole. Earth Ground Stations (EGSs) and GEO satellites are effectively represented in the ECEF coordinate frame. $R_e$ is denoted as the radius of earth\footnotemark[1]. We can calculate the coordinates $(x_{ECEF},y_{ECEF},z_{ECEF})$ of an EGS or GEO satellite in the ECEF frame using its $(L,B,H)$, where $L$, $B$ and $H$ are its longitude, latitude and height, then:
\footnotetext [1] {In this article, we consider the earth and the moon as uniform spheres.}
\begin{equation}
\begin{array}{l}
{x_{ECEF}} = ({R_e} + H)\cos B\cos L\\
{y_{ECEF}} = ({R_e} + H)\cos B\sin L\\
{z_{ECEF}} = ({R_e} + H)\sin B. \label{13}
\end{array}
\end{equation}

In the ECI frame, the $x$-$y$ plane and $z$-axis are the same as those of the ECEF frame, the $x$-axis is fixed and points towards the vernal equinox of the year 2000. The moon revolves around the earth, however, the ECI coordinate frame is inertial and does not rotate with the earth, allowing for a unified representation of the Earth-Moon coordinate system. Thus, we can effectively describe the positions and motions within the Earth-Moon system by utilizing the ECI frame. Greenwich Mean Sidereal Time (GMST) \cite{a49,a50} is an important astronomical time scale used to describe the angle of the earth's rotation, GMST can be calculated in the following way: $GMST = 18.697374558 + 24.06570982441908 \times D$, where $D$ is the Julian day beginning with J2000, $18.697374558$ is the GMST of J2000, and $24.06570982441908$ is the angle of the earth's rotation per solar day. 

Therefore, the coordinate $(x_{ECEF},y_{ECEF},z_{ECEF})$ of the EGSs and GEO satellites in the ECEF frame can be transformed into the coordinate $(x_{ECI},y_{ECI},z_{ECI})$ in the ECI coordinate system by a rotation matrix ${\boldsymbol{R} _z}({ -GMST})$,
${\boldsymbol{R} _x}({ \cdot})$ and ${\boldsymbol{R} _z}({ \cdot})$ are the rotation matrices\cite{a50}:
\begin{equation}
\left[ {\begin{array}{*{20}{c}}
{{x_{ECI}}}\\
{{y_{ECI}}}\\
{{z_{ECI}}}
\end{array}} \right] = {\boldsymbol{R} _z}({ -GMST}) \cdot \left[ {\begin{array}{*{20}{c}}
{{x_{ECEF}}}\\
{{y_{ECEF}}}\\
{{z_{ECEF}}}
\end{array}} \right].\label{16}
\end{equation}

\subsubsection{LCE Transform to ECI}

\begin{figure}[t]
\centering
\includegraphics[width=0.47\textwidth]{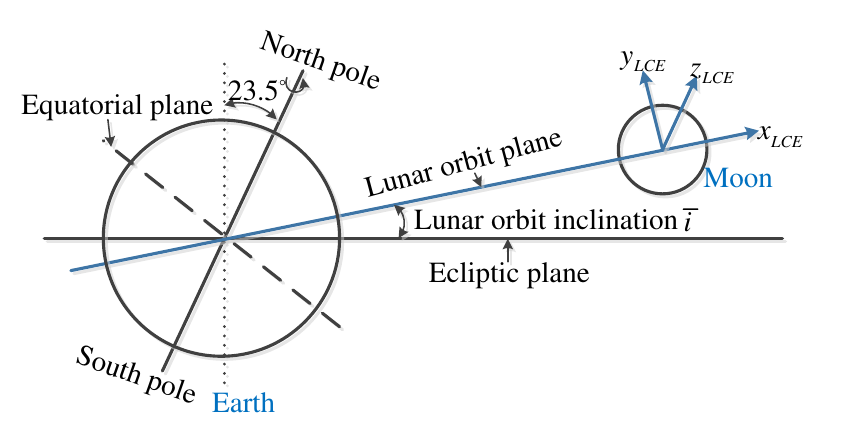}
\captionsetup{font=small,labelsep=period, singlelinecheck=false}
\captionsetup{labelformat=modified, justification=raggedright}
\caption{\small The LCE and ECI coordinate frame.} \label{fig6}
\end{figure}
The lunar network coordinates will be converted from the LCE coordinate frames to the ECI frames. Before we introduce the coordinate transformation of the LCE frame to the ECI frame, we will first introduce the moon's orbit around the earth. In the Geocentric Ecliptic Coordinate (GEC) system of epoch J2000, the origin is in the center of the earth, the $x$-axis points in the direction of the vernal equinox, the $z$-axis is perpendicular to the ecliptic plane, and the $x$-$y$-$z$ axis is a right-handed spiral, respectively. As for the six orbital elements of the moon orbiting the earth $(\bar a, \bar e, \bar i, \bar \Omega, \bar \omega, \bar M)$, please refer to formula $(1.97)$ in \cite{a50}.

Fig. \ref{fig6} shows the LCE and ECI coordinate frame, as can be seen from the figure, the earth's rotation axis and the ecliptic plane have an angle of 23.5°. By solving the Kepler equation, we can obtain the position of the moon center $(x_{orb}^m,y_{orb}^m,z_{orb}^m)$ in the orbital plane, and obtain the coordinates $(x_{GEC}^m,y_{GEC}^m,z_{GEC}^m)$ of the moon center in the GEC system through the rotation matrix $R1$, and then obtain the coordinates $(x_{ECI}^m, y_{ECI}^m, z_{ECI}^m)$ of the moon center in the ECI coordinate frame through the rotation matrix $R2$:
\begin{equation}
\left[ {\begin{array}{*{20}{c}}
{x_{ECI}^m}\\
{y_{ECI}^m}\\
{z_{ECI}^m}
\end{array}} \right] = R2 \cdot R1 \cdot \left[ {\begin{array}{*{20}{c}}
{x_{orb}^m}\\
{y_{orb}^m}\\
{z_{orb}^m}
\end{array}} \right], \label{18}
\end{equation}
where $R1 = {\boldsymbol{R} _z}({\bar \Omega}) \cdot {\boldsymbol{R} _x}(\bar i) \cdot {\boldsymbol{R} _z}(\bar w)$, $R2 ={\boldsymbol{R} _x}(23.5^\circ)$.   

As for the coordinate $(x_{LCE}^o, y_{LCE}^o, z_{LCE}^o)$ of a Halo orbit or an ordinary lunar orbit in the LCE frame, we can first move the origin of the coordinate from the moon's center to earth's center to get the coordinates $(x_{orb}^o, y_{orb}^o, z_{orb}^o)$ in the moon orbital plane: $(x_{orb}^o,y_{orb}^o,z_{orb}^o) = (x_{LCE}^o,y_{LCE}^o,z_{LCE}^o) + ( \sqrt {{{(x_{orb}^m)}^2} + {{(y_{orb}^m)}^2} + {{(z_{orb}^m)}^2}},0,0)$,
then we can get the coordinate $(x_{LCE}^o, y_{LCE}^o, z_{LCE}^o)$ converted to the ECI frame by the rotation matrices:
\begin{equation}
\left[ {\begin{array}{*{20}{c}}
{x_{ECI}^o}\\
{y_{ECI}^o}\\
{z_{ECI}^o}
\end{array}} \right] = R2 \cdot R1 \cdot \left[ {\begin{array}{*{20}{c}}
{x_{orb}^o}\\
{y_{orb}^o}\\
{z_{orb}^o}
\end{array}} \right]. \label{22}
\end{equation}
\vspace{0.10cm}
\subsection{Age of Information}
In this paper, AoI is used to characterize the information timeliness of the system, the channel is considered unreliable due to the significant path loss. Assume a data packet generated by the source node is relayed to the EGS via an $n$-hop path $P_i$. Let $p_j$ denote the transmission failure probability of the $j$-th hop relay link, which can be derived from the channel's bit error rate (BER). Then, the probability of this packet successfully reaching the destination node is $p^s = \prod\limits_{j = 1}^n {(1 - {p_j})}$. Define $T(P_i)=d_i - g_i$ as the total transmission AoI of packet $f_i$ which is transferred via path $P_i$, where $d_i$ and $g_i$ are arrival time and generation time, respectively. As for $T(P_i)$ of $n$-hop path, we mainly consider two delays: propagation delay $T_{pro}$ and transmission delay $T_{tran}$, ${T(P_i)} = {T_{pro}} + {T_{tran}}$. We set unit distance of unit AoI is $D_{unit}$, $T_{pro}$ is determined by the total distance of the $n$-hop path $D_{total}$, $T_{pro}=D_{total}/D_{unit}$. The transmission delay $T_{tran}$ can be calculated by $T_{tran} = F_i/T_{rate}/(D_{unit}/c)$, where $c=3*10^5$ km/s is the speed of light, $F_i$ is the packet size of $f_i$, and $T_{rate}$ is the transmission rate of relay or source node.
\begin{figure}[t]
\centering
\includegraphics[width=0.47\textwidth]{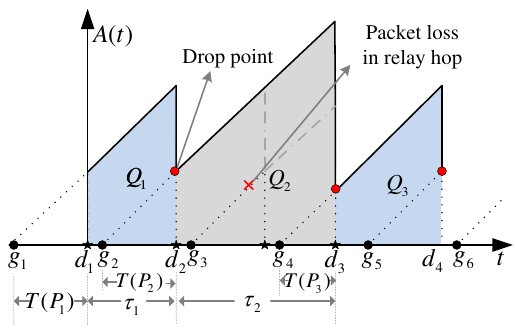}
\captionsetup{font=small,labelsep=period, singlelinecheck=false}
\captionsetup{labelformat=modified, justification=raggedright}
\caption{\small AoI model.} \label{fig2}
\end{figure}

The AoI $A(t)$ drops to $T(P_i)=d_i - g_i$ when the last generated update arrives at the destination node successfully. Otherwise, $A(t)$ will increase linearly, as shown in Fig. \ref{fig2}, the evolution of $A(t)$ in a multiple hop network is:
\begin{equation}
A(t + 1) = \left\{ {\begin{array}{*{20}{c}}
{A(t) + 1, \ {\rm{ if }}\ \varpi {\rm{(t) = 0}}}\\
{t - g(i),\ {\rm{ if }}\ \varpi {\rm{(t) = 1}}},
\end{array}} \right. \label{5}
\end{equation}
where ${\varpi {\rm{(t)}}}$ is the binary indicator factor to indicate the success (${\varpi {\rm{(t)}}}=1$) or failure (${\varpi {\rm{(t)}}}=0$) of a packet transmission through $n$-hop path $P$. Therefore, $\Pr \{ \varpi {\rm{(t) = 1}}\} = p^s$; otherwise,  $\Pr \{ \varpi {\rm{(t) = 0}}\} =1- p^s$. 

We assume that each $s_m$ generates update packets at equal intervals $\sigma=g_i-g_{i+1}$, then the arrival interval of two successive packets at the destination node is ${\tau _i} = {d_{i + 1}} - {d_i}$, $t=1, 2, 3...$ is a sequence of time slots. For simplicity, our formula below will omit $t$. The number of packets generated in the observation time $(0, \mathcal{T})$ is $K(\mathcal{T})$, we define effective arrival rate is the number of AoI drops per unit time, then $\mu  = {[\sum\limits_{i = 1}^{K({\mathcal T})} {{\varpi _i}} }]/{{\mathcal T}}$, where $K(\mathcal T) ={\mathcal T}/{\sigma }$, $\mu {\mathcal T}$ indicates the total number of packets arrived. The average AoI during observation time $(0, \mathcal T)$ can be calculated as:
\begin{equation}
\Delta \bar A = \frac{1}{\mathcal T} \sum_{i = 1}^{\mu \mathcal T} Q_i = \frac{1}{\mathcal T} \sum_{i = 1}^{\mu \mathcal T} \left[\frac{1}{2} \left(\tau_i + T(P_i)\right)^2 - \frac{1}{2} T(P_i)^2\right],\label{3}
\end{equation}
where $Q_i$ is the area of $i$-th quadrilateral in Fig. \ref{fig2}; $\{ {\tau _1},{\tau _2},...,{\tau _{\mu {\mathcal T}}}\}  \in \Gamma_1 ,\{ T({P_1}),T({P_2}),...,T({P_{\mu {\mathcal T}}})\}  \in \Gamma_2  $, $\Gamma_1$ and $\Gamma_2$ are the set of packet arrival intervals and transmission delays in the system, respectively. Assuming that the average arrival interval of the $\mu \mathcal T$ arrived packets is $\bar \tau$, $\bar \tau  = 1/\mu$, then:
\begin{align}
\Delta \bar A &= \frac{1}{{\mathcal T}}\sum\limits_{i = 1}^{\mu {\mathcal T}} {[\frac{1}{2}{\tau _i}^2 + {\tau _i} \cdot T({P_i})]} = \frac{1}{{\mathcal T}} \cdot \mu {\mathcal T} \cdot \frac{1}{2}{{\bar \tau }^2} \notag \\  
&+ \frac{1}{{\mathcal T}}\sum\limits_{i = 1}^{\mu {\mathcal T}} {\bar \tau   \cdot T({P_i})} = \frac{\bar \tau}{{2}} + \frac{1}{{{\mathcal T} \cdot \mu }}\sum\limits_{i = 1}^{\mu {\rm T}} { T({P_i})} \notag \\  &=  \frac{\sigma }{{2\prod\limits_{j = 1}^n {(1 - {p_j})} }} + \bar T({P_i}), \label{7}
\end{align} 
where $\bar \tau  = {\mathds{E}}[{1}/{\mu}] = {\mathds{E}}[{\mathcal {T}/[\sum\limits_{i = 1}^{K({\mathcal T})} {{\varpi _i}} }]] ={\mathds{E}}[{\mathcal T}/({ K({\mathcal T}) \cdot {p^s}}) ]  =\sigma / \prod\limits_{j = 1}^n {(1 - {p_j})} $, $\bar T({P_i}) = \frac{1}{{\mu {\mathcal T} }}\sum\limits_{i = 1}^{\mu {\mathcal T}} {T({P_i})} $. ${\mathds{E}}\{  \cdot \} $ is the expectation with respect to the system randomness (i.e., channel variation and packet arrival processes). We denote the AoI of source $s_m$ going through path $P_i$ in slot $t$ at EGS is $\Delta \bar A_m^{P_i}[t]$. A packet going through different paths $P_i$ may have different AoIs at EGSs, because we aim to improve the timeliness of the system, so we assume that each packet chooses the path minimizing the AoI at EGS, then the final AoI of $s_m$ in slot $t$ at EGS is $A_m[t]=min{ \{\Delta \bar A_m^{P_1}[t], \Delta \bar A_m^{P_2}[t], ...,\Delta \bar A_m^{P_i}[t]\}}$. 


\begin{figure*}[t]
\centering
\includegraphics[width=0.9\textwidth]{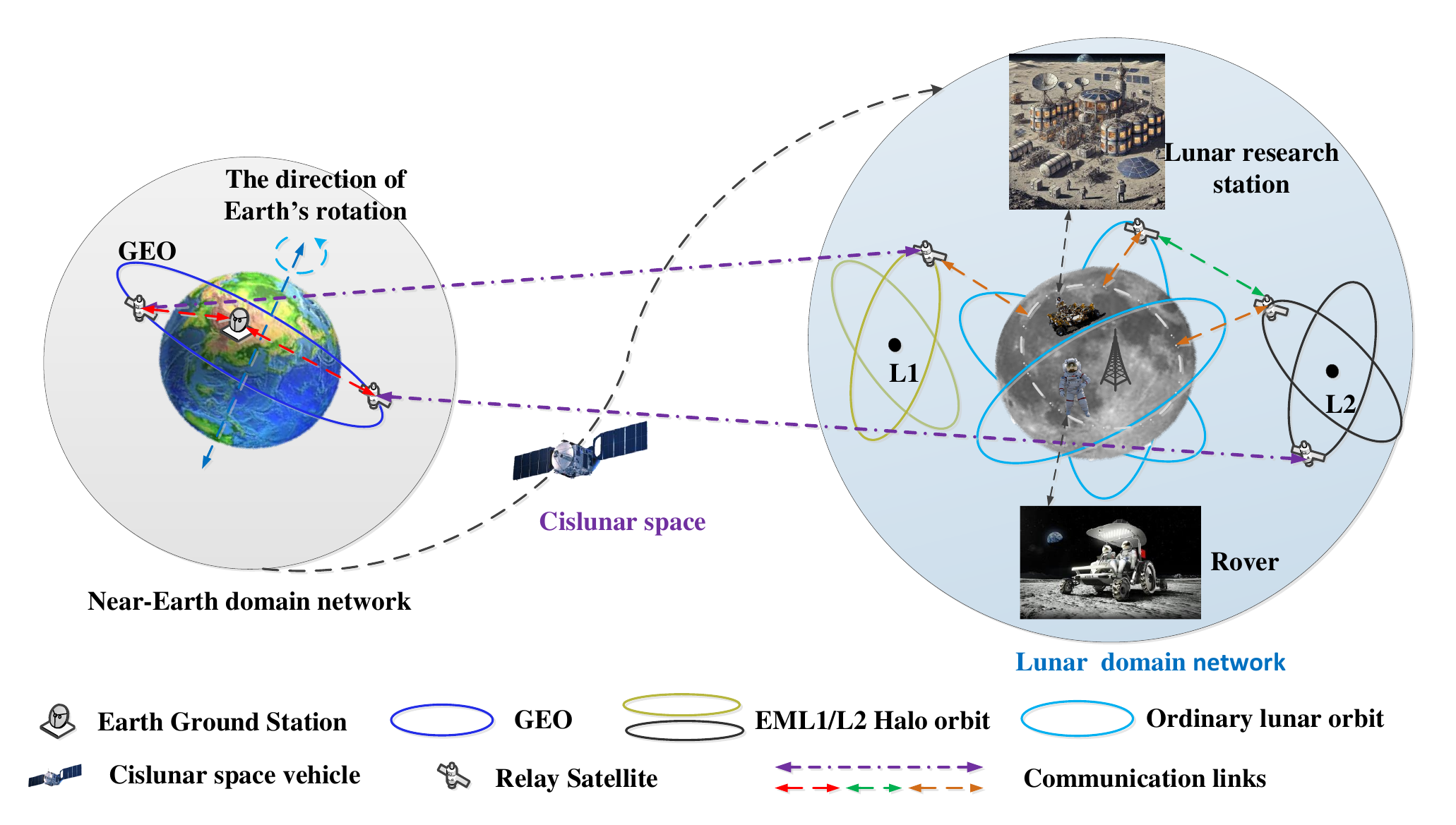}
\captionsetup{font=small,labelsep=period, singlelinecheck=false} 
\captionsetup{labelformat=modified, justification=raggedright} 
\caption{\small The proposed Earth-Moon heterogeneous relay system.}
\label{fig1}
\end{figure*}

\section{Problem Formulation}
\subsection{System Model}
We consider the earth's rotation and the moon's revolution to design a combined Earth-Moon hybrid relay satellite constellation to support real-time lunar exploration missions as shown in Fig. \ref{fig1}, which consists of a near-earth domain network, cislunar space, and a lunar domain network. The near-earth domain network includes $N_{GEO}$ GEO satellites and $N_{EGS}$ EGSs. The cislunar space contains several spacecrafts. The lunar domain network consists of $M$ lunar source nodes (i.e., rovers, astronauts, and lunar research stations), $N_{L1}$ EML1 Halo orbit satellites, $N_{ord}$ ordinary lunar orbit satellites (i.e., elliptical orbit or circular orbit), and $N_{L2}$ EML2 Halo orbit satellites, respectively. Besides, Fig. \ref{fig1} depicts the relative positions of EML1, EML2 and moon. The EML1 and EML2 Halo orbit satellites provide superior coverage for the Lunar Near Side (LNS) and LFS, respectively, and their links to the earth remain unblocked by the moon. In particular, Halo orbits are divided into southern and northern family Halo orbits, whose unique characteristics are described in more detail later. EML1/L2 orbit satellites can not cover the whole lunar surface, ordinary lunar orbit satellites can bridge the coverage gaps of the EMLP orbit satellites, ensuring comprehensive coverage. The relay constellation configuration structure is denoted by \{$N_{GEO}$, $N_{L1}$, $N_{ord}$, $N_{L2}$\}.

Consequently, a novel Earth-Moon Relay Communication Network  Architecture (EMRCNA) is proposed based on the distances from various orbits to Earth. EMRCNA includes four groups of relay satellites and one group of EGSs. Respectively, the first group consists of $N_{L2}$ EML2 Halo orbit satellites, the second group includes $N_{ord}$ ordinary lunar orbit satellites, the third group is composed of $N_{L1}$ EML1 Halo orbit satellites, the fourth group contains $N_{GEO}$ GEO satellites, and the fifth group consists of several EGSs from a specific country. Data packets generated by lunar probes and astronauts on the moon can be relayed via the lunar domain network to earth domain EGSs, with lower-numbered group satellites transmitting to higher-numbered group satellites when a Line-of-Sight (LoS) link is available, no data transmission occurs within the same group. Notably, a group is not restricted to forwarding data only to the next sequential group, packets can be transmitted directly to any higher-numbered group, provided an LoS link exists. The optimal transmission path for a data packet is determined by the path that maximizes system freshness. EMRCNA ensures a systematic and reliable transfer of information, minimizing unnecessary data relays from LNS to LFS and back to earth.

\subsubsection{Orbital Model}
\begin{figure}[t] 
\centering
\includegraphics[width=0.47\textwidth]{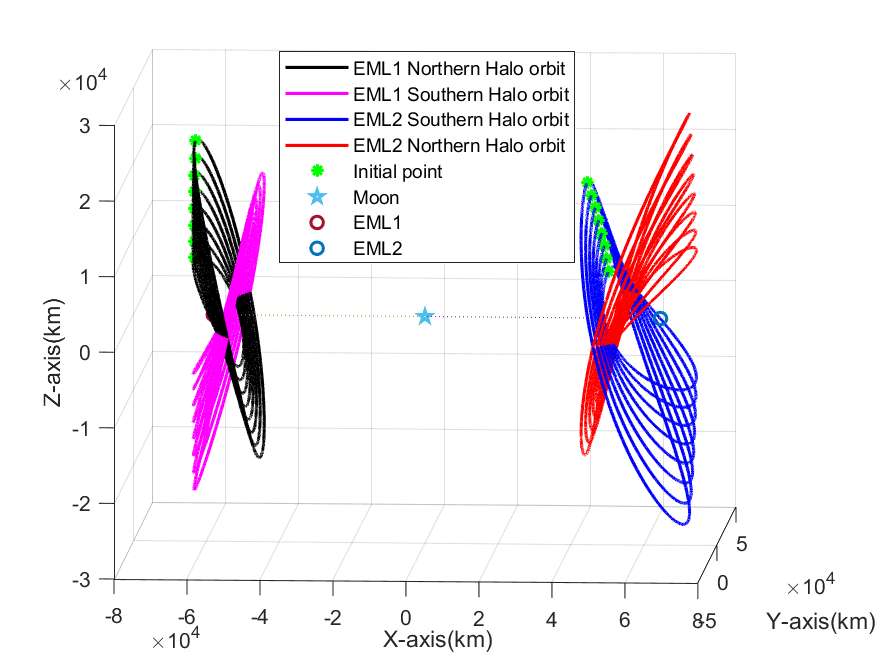}
\captionsetup{font=small,labelsep=period, singlelinecheck=false}
\captionsetup{labelformat=modified, justification=raggedright}
\caption{\small EML1/EML2 southern and northern family Halo orbits.} \label{fig3}
\end{figure}

The characteristics and parameters of Halo orbits and ordinary lunar orbits are considered. In the Earth-Moon circular-restricted three-body problem \cite{a42}, there are three collinear (L1, L2, L3) and two triangular (L4, L5) Libration points, where the deployed satellites will remain stationary because of their force balance. L1 is located between earth and moon along the line connecting them, while L2 is situated on the extension of the Earth-Moon line. Satellites deployed at the EML1/L2 point provide excellent coverage of the LNS/LFS, respectively. In the LCE coordinate system, we define the plane of the moon's orbit around the earth as the $x$-$y$ plane, where the direction of earth towards to moon is the $x$-axis direction, the $y$-axis is the moon's revolution direction, and $z$ and $x$-$y$ are right-handed spiral rule, respectively. The EML1/L2 Halo orbit is divided into the EML1/L2 northern and southern family Halo orbits, where the northern Halo orbits and the southern Halo orbit of the same amplitudes are symmetric about the $x$-$y$ plane. More characteristics of individual Halo orbits can be referred to in our previous work \cite{a43, a44}. Fig. \ref{fig3} shows the EML1/L2 northern and southern Halo orbits with different $z$-amplitudes $A_z$. It is worth mentioning that given the parameters L1 or L2, southern or northern family Halo orbits, and $A_z$ can uniquely determine a Halo orbit. Therefore, a Halo orbit can be determined by the parameter set $\Psi $, which includes $\{L, F, A_z\}$ of three parameters. In particular, $L$ is the indicator of EMLP, where $L=1$ ($L=2$) is EML1 (EML2), respectively. $F$ is the indicator of the north and south, where $F=n$ ($F=s$) is the north (south) family Halo orbits. 
\footnotetext [2] {The Prime Meridian line is similar to the earth's Prime Meridian line in that it refers to the center line on the LNS.}

\begin{figure}[t]
\centering
\includegraphics[width=0.47\textwidth]{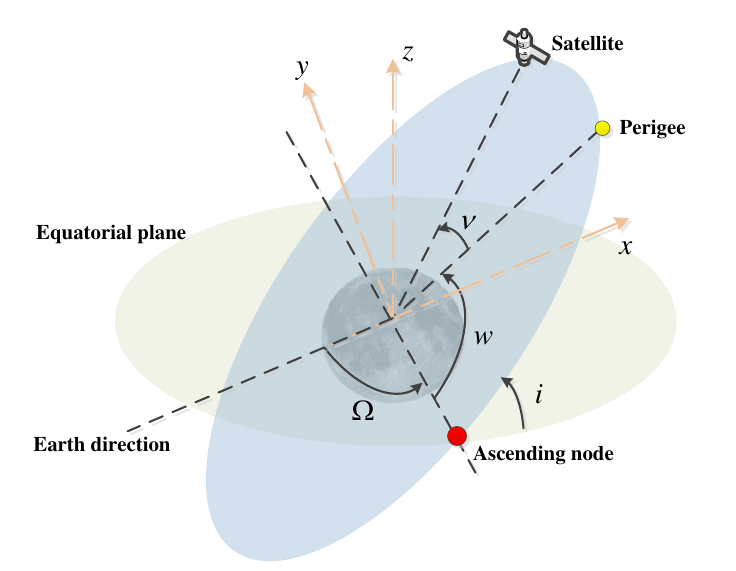}
\captionsetup{font=small,labelsep=period, singlelinecheck=false}
\captionsetup{labelformat=modified, justification=raggedright}
\caption{\small Ordinary lunar orbit.} \label{fig4}
\end{figure}

In the LCE coordinate system, the position of an ordinary lunar orbit satellite can be determined by the following parameters $\Phi =\{a, e, i, \Omega, w, \nu\}$ \cite{a45}, as shown in Fig. \ref{fig4}: Semi Major Axis ($a$), it is half the major axis (radius) of the elliptical (circular) orbit, respectively;
Eccentricity ($e$), $0 < e < 1$ ($e=0$), it indicates an elliptical (circular) orbit, respectively; Inclination ($i$), it refers to the angle measured counterclockwise from the lunar equator to the orbital plane at the right ascending node, ${0^ \circ } \le i \le {180^ \circ }$; Right Ascension of Ascending Node ($\Omega$), it measures counterclockwise from the point at which the moon's Prime Meridian\footnotemark[2] intersects the equator to the ascending node, when viewed over the north pole, in the range $[0^\circ, 360^\circ]$; Argument of the Perigee ($w$), it measures counterclockwise along the orbit from the ascending node to the perigee. The value is $0^\circ$ when the orbit is circular; True Anomaly ($\nu$), it is the geocentric angle between the perigee direction and the satellite direction. The position of each satellite can be expressed by a function of these angles, as follows:
\begin{equation}
\left( {\begin{array}{*{20}{c}}
{\!\!{x^s}\!\!}\\
{\!\!{y^s}\!\!}\\
{\!\!{z^s}\!\!}
\end{array}} \right) = a\left( {\begin{array}{*{20}{c}}
{\!\!\! \cos (w + \nu )\cos \Omega  - \sin (w + \nu )\cos i\sin \Omega \!\!\!}\\
{\!\!\! \cos (w + \nu )\sin \Omega  - \sin (w + \nu )\cos i\cos \Omega \!\!\! }\\
{\!\!\! \sin (w + \nu )\cos i \!\!}
\end{array}} \right)\!. \label{8}
\end{equation}

\subsubsection{Coverage Model}
Similar to earth's latitude and longitude coordinates, we use latitude and longitude to represent the position of observation points on the moon. The moon's Prime Meridian is designated as the $0^\circ$ longitude. The lunar eastern hemisphere has longitudes ranging from $[0^\circ, 180^\circ \rm{E}]$, and the western hemisphere ranges from $[-180^\circ \rm{W}, 0^\circ]$. The $0^\circ$ latitude plane is the equatorial plane, with the northern hemisphere having latitudes ranging from $[0^\circ, 90^\circ \rm{N}]$ and the southern hemisphere ranging from $[-90^\circ \rm{S}, 0^\circ]$. Notably, the LNS has a longitude range of $[-90^\circ \rm{W}, 90^\circ \rm{E}]$, and the LFS has a longitude range of $[90^\circ \rm{E}, 180^\circ \rm{E}]$ and $[-180^\circ \rm{W}, -90^\circ \rm{W}]$.

If the lunar observation points are sampled at equal latitude and longitude intervals, they will concentrate in the north and south regions of the moon. To avoid this situation, we use the Fibonacci lattice \cite{a27, a41} method to generate $M$ lunar observation points uniformly to evaluate the coverage of the designed constellation on the lunar surface. 
The radius of the moon is $R_m$, then coordinates of the $m$-th observation point $s_m$ in the LCE Coordinate system is $(x_s^m, y_s^m, z_s^m)$:
\begin{equation}
\begin{array}{l}
z_s^m = {R_m} \cdot (2m - 1)/M - 1,\\
x_s^m = {R_m} \cdot\sqrt {1 - {{(z_s^m)}^2}}  \cdot \cos (2\pi m\phi ),\\
y_s^m = {R_m} \cdot\sqrt {1 - {{(z_s^m)}^2}}  \cdot \sin (2\pi m\phi ),\label{system1}
\end{array}
\end{equation}
where $\phi$ is taken to obey the golden ratio under the requirement of Fibonacci lattice, i.e. $\phi=(\sqrt{5}-1)/2$. Notably, if the target areas for our objectives differ, such as optimizing AoI over the lunar surface and coverage in the South Pole, using the Fibonacci lattice method to generate $M$ points across the surface may place only $0.1M$ in the South Pole. To address this, we ensure equal observation points for all metrics by generating additional points to maintain $M$ points specifically in the South Pole region. 

The grid point coverage analysis method \cite{a43} is used to analyze the coverage states of the constellation on the lunar surface. Coverage can be achieved as long as $\theta$ is greater than the given observation elevation angle constraint $\theta_{c}$, where $\theta$ is the observation angle of the observation points to the satellites. 
${{\vec r}_u} $ represents the vector from the center of the moon to the observation point $s_m$ on the lunar surface, and ${{\vec r}_{us}}$ is the vector from $s_m$ to the relay satellite $r_k$ respectively, then $\theta$ can be calculated according to the following formula: 
\begin{equation}
\theta  = {90^ \circ } - \arccos (\frac{{{{\vec r}_u} \cdot {{\vec r}_{us}}}}{{\left| {{{\vec r}_u}} \right| \cdot \left| {{{\vec r}_{us}}} \right|}}). \label{2}
\end{equation}

By calculating the coverage states of every observation point on moon, we can determine the coverage states of the lunar surface by the relay network. We define $cov$ as the percentage of continuous one-fold coverage of the designed constellation on the target detection area by calculating the coverage factor $z_k^m, k \in \{ 1,2,...,N{_{L2}}+N{_{ord}}+N{_{L1}} \}$ one by one:
\begin{equation}
{\mathop{cov}}  = \frac{{\sum\nolimits_{{s_m} \in S} {{z^m}} }}{M},{\rm{ }}m \in \{ 1,2,3...M\},\label{3}
\end{equation}
where $S$ indicates the lunar target detection area, ${z^m}=1$ if $s_m$ is covered by any relay satellite $r_k$, that is ${z^m}=1$ if $\exists \ z_k^m = 1{\rm{ }}, \forall \ k$; ${z^n}=0$, otherwise. In addition, Table I lists the main notations in this paper.
\begin{table}[t]
\setlength{\abovecaptionskip}{1pt}
\renewcommand\arraystretch{1.5}
\centering
\footnotesize
\begin{threeparttable}
\vspace{-0.5em}
\caption{\small Notations}
\begin{tabular}{l|l} 
\toprule
\multicolumn{1}{c|}{\textbf{Notation}} & \multicolumn{1}{c}{\textbf{Definition}} \\
\hline 
$M$ & The total number of observation points on the moon\\
${s_m}$ & A source node on the moon, where $m = 1, \cdots ,M$\\
$N$ & The total number of relay satellites in the network\\
${r_k}$ & Relay satellite in the network, where $k = \{ 1,2\cdots N\} $\\
${R_m}$ & The radius of the moon \\
${R_e}$ & The radius of the earth \\
$\theta $ & The angle between ${s_m}$ and ${r_k}$ \\
${z_k^m}$ & Indicator factor of the coverage of ${r_k}$ over ${s_m}$ \\
$p_j$  & The transmission failure probability of the $j$-th hop relay link\\
$p^s$ & The success probability of a packet reaching EGS\\
$T(P_i)$ & The total transmission delay of path $P_i$\\
$\sigma$  & The interval of generated update\\
$\tau_i$  & The arrival interval of two successive packets at EGS \\
$A_z$ & The $z$-axis amplitude of Halo orbit\\
$D_{unit}$ & The unit distance of unit AoI\\
${T_{ord}}$ & The period of ordinary lunar orbit \\
${T_{halo}}$ & The period of Halo orbit \\
$\bar A $  & The average per-device AoI \\
$\Psi $ & The parameter set of a Halo orbit satellite\\
$\Phi$ & The parameters set of an ordinary lunar orbit satellite\\
$D_i$ & The crowding-distance of the $i$-th solution\\
$f_m^{obj}$ & The value of $m$-th objective for $i$-th individual\\
\bottomrule
\end{tabular}
\end{threeparttable}
\end{table}

\subsection{Problem Formulation}
In this article, we aim to design an age and coverage optimal Earth-Moon relay satellite communication constellation with a minimum number of satellites to meet the data freshness requirements of future lunar missions and related activities. Such constellation design is to leverage the unique characteristics of EML1/L2 Halo orbits. The EML1 (EML2) satellite provides good coverage of the LNS (LFS), respectively. We analyze the coverage status of the satellites in the Halo orbits at EML1/L2 and find that $N_{L1}$ EML1 and $N_{L2}$ EML2 satellites cannot fully cover the entire lunar surface, due to the symmetry of EML1/L2, so we introduce $N_{ord}$ circular lunar orbit satellites to achieve complementary coverage, which enables real-time full coverage of the specific lunar surface during lunar exploration process. However, despite 100\% coverage of the lunar surface by lunar satellites, the earth cannot communicate with the lunar network in real-time due to visibility of geometry based communication window issues caused by earth's rotation and the moon's revolution. GEO orbital period matches the earth's rotation period and GEO satellites provide great coverage of the earth's surface. Therefore, we introduce $N_{GEO}$ GEO satellites to facilitate a relay network design between EGSs and the lunar domain network, optimizing the overall system's average AoI to enhance information timeliness.

We define $T_{halo}$ and $T_{ord}$ as the period of the Halo orbit and ordinary lunar orbit respectively. To prevent the period of the designed lunar relay constellation from being too large, all Halo orbits should have the same $A_z$, and all ordinary lunar orbits have the same semi major axis $a$. The following condition needs to be met:
\begin{equation}
mod(Y \cdot T_{halo}, T_{ord})=0, \label{9}
\end{equation}
where $Y$ is a positive integer. Here, we assume that $1 \le Y \le 2$, then the period of the lunar relay constellation can be represented as $Y \cdot T_{halo}$. After $A_z$ is determined, we can get $T_{ord}$ according to the period constraint formula mentioned above. Then, we can calculate ordinary lunar orbits' semi major axis $a$ through the following formula:
\begin{equation}
a = \frac{{\sqrt[3]{{GM*T_{ord}^2}}}}{{4{\pi ^2}}},\label{10}
\end{equation}
where the lunar gravitational constant is $GM=4.903 \times 10^{12} m^3/s^2$.

Additionally, we aim to achieve the aforementioned real-time communication with the minimum number of satellites $N=N_{GEO}+N_{L1}+N_{ord}+N_{L2}$. We define the average per-device AoI at EGSs during observation time $(0, \mathcal T)$ as $\bar A$:
\begin{equation}
\bar A  \buildrel \Delta \over = \mathop {\lim {\rm{ sup}}}\limits_{\mathcal T \to \infty } \frac{1}{{\mathcal T \cdot M}}{\mathds{E}}\left[ {\sum\limits_{t = 1}^{\mathcal T} {\sum\limits_{m = 1}^M {{A_m}[t]} } } \right].\label{11}
\end{equation}

We assume the parameters of $i$-th GEO satellite, $j$-th Halo orbit satellite and the $k$ ordinary lunar orbit satellite are $\xi_i $, $ \Psi _j= \{L_j, F_j, A_z^j\}$, $\Phi_k=\{a_k, e_k, i_k, \Omega_k, w_k, \nu_k\}$ respectively. By these definitions, we intend to find the optimal relay satellite constellation with parameters $\xi_i, \Psi_j, \Phi_k$ that minimize the average per-device AoI $\bar A$ and maximize the coverage of the target area $cov$ with a minimum number of satellites $N$, then a multiple objective optimization problem is formulated as follows: 
\begin{equation}
\begin{split}
\begin{array}{l}
\begin{array}{*{20}{c}}
{\qquad\mathop {\min }\limits_{\xi_i, \Psi_j, \Phi_k } }&{N, \bar A, - cov}
\end{array}\\
\begin{aligned}
s.t. \quad C1&: a_{k1}=a_{k2}, k1,k2 \in \{1,2,...,N_{ord}\},  \vspace{0.1cm}\\
C2&:  e_k, w_k = 0,   \vspace{0.1cm}\\
C3&: {0^ \circ } \le i_k \le {180^ \circ },   \vspace{0.1cm}\\
C4&: {0^ \circ } \le \Omega_k, \nu\_k \le {360^ \circ } \vspace{0.1cm},\\
C5&: {a_k}(1 - {e_k}) - {R_m} > 100, k \in \{1,2,...,N_{ord} \}, \vspace{0.1cm}\\
C6&: {0^ \circ }\le \xi_i \le  {180^ \circ } ,  i \in \{1,2,...,N_{GEO} \}, \vspace{0.1cm}\\
C7&: A_z^{j1}=A_z^{j2}, j1,j2 \in \{1,2,...,N_{L1} +N_{L2}\},  \vspace{0.1cm}\\
C8&: \bmod (Y \cdot {T_{halo}},{T_{ord}}) = 0, \\
C9&: N=N_{GEO}+N_{L1}+N_{ord}+N_{L2}, \label{12}\\
\end{aligned}
\end{array}
\end{split}
\end{equation}
where $C1$ is to ensure the period of all ordinary lunar orbit satellites is the same; $C2$-$C4$ are the parameter range constraints of ordinary lunar orbits; $C5$ is the perigee height constraint, if the orbit's perigee is below 100 km, the orbit is unstable \cite{a46}; $C6$ is the GEO longitude constraint; $C7$ is the Halo amplitude constraint, $C8$ is the period constraint of Halo and ordinary lunar orbit, and $C9$ is calculation formula for total number of satellites, respectively. 

\section{Solution and Algorithms}
In this section, we detail the approach to solving the constellation design problem, as formulated in equation \eqref{12}, using the NSGA-II algorithm, which is a well-established method for addressing multi-objective optimization problems, utilizing fast nondominated sorting and crowding distance assignment to ensure diversity in the Pareto front \cite{a51,a52}. 
Given the complexity of the objective functions, we approach the decomposition of problem \eqref{12} from the perspective of constellation architecture. Once the architecture is defined, i.e., the number of satellites allocated to GEO ($N_{GEO}$), EML1/L2 Halo orbits ($N_{L1}$, $N_{L2}$), and ordinary lunar orbits ($N_{ord}$), the total number of satellites $N$ is determined. Consequently, $N$ can be decoupled from problem \eqref{12} by simulating various network configurations. Subsequently, the problem is solved considering the two objectives: the average per-device AoI and coverage. Therefore, the optimization problem under a fixed $N$ then becomes:
\begin{equation}
\begin{split}
\begin{array}{l}
\begin{array}{*{20}{c}}
{\qquad\mathop {\min }\limits_{\xi_i, \Psi_j, \Phi_k } }&{\bar A, - cov}
\end{array}\\
\begin{aligned}
s.t. \quad C1,C2,C3,C4,C5,C6,C7,C8 \ in \ \eqref{12}. \label{23}
\vspace{0.1cm}\\
\end{aligned}
\end{array}
\end{split}
\end{equation}

By utilizing a customer-designed NSGA-II, we can effectively explore AoI and coverage performance under various constellation configurations. NSGA-II identifies optimal or near-optimal solutions that balance AoI and coverage objectives, leading to the determination of optimal constellation parameters for a fixed $N$. \textbf{Algorithm 1} presents the preudo-code of NSGA-II based constellation design, with key steps outlined below:

\textbullet \enspace \textit{Population initialization}: For a fixed constellation architecture, specified by $\{N_{GEO},N_{L1},N_{ord},N_{L2}\}$, the parameters for each individual ${ \mathds{I}}= \{ {\xi _i},{\Psi _{j1}},{\Phi _k},{\Psi _{j2}}\},i \in \{ 1,2...,{N_{GEO}}\} ,j1 \in \{ 1,2...,{N_{L1}}\},k \in \{ 1,2...,{N_{ord}}\}, j2 \in \{ 1,2...,{N_{L2}}\} $ are initialized according to the value ranges defined by constraints $C2$-$C4$ and $C6$ in problem \eqref{23}. For a determined constellation, certain parameters are preset; For example, i.e., if one satellite is positioned in an EML2 southern Halo orbit, then $ \Psi = \{2, s, A_z\}$. 
\begin{algorithm}[t]
\small
\renewcommand{\baselinestretch}{0.8}
\caption{Joint Age and Coverage Optimization}
\LinesNumbered
\KwIn{network configuration $N_{GEO}, N_{L1}, N_{ord}, N_{L2}$, population size $\Re$, number of generations $G_{max}$, mutation rate, crossover rate}
\KwOut{Pareto front $\Im^{1}$ in generation $G_{max}$}
generate $\Re$ individuals ${ \mathds{I}}=\{\xi, \Psi, \Phi\}$ randomly as the first parent generation $P(1)$\;
set offspring $Q_1=\emptyset$\;
\For{$G \leftarrow 1 : G_{max}$} {
    ${R_G} = {P_G} \cup {Q_G}$\;
    calculate $\bar{A}$ and $cov$ based on formula \eqref{3} and \eqref{11}, and check the constraints $C5$ in problem \eqref{23} for each individual in $R_G$\;
    $\Im$ = fast nondominated sort ($R_G$) based on \textbf{Algorithm 2}\;
    $P_{G+1}=\emptyset$ and $l=1$\;
    \While{$\left| {P_{G+1}} \right| + \left| {\Im^l} \right| \le \Re$} {
        calculate crowding-distance of (${\Im^l}$)\;
        $P_{G+1} = P_{G+1} \cup {\Im^l}$\;
        $l = l + 1$\;
    }
    sort(${\Im^l}$, ${\prec_n}$)\;
    $P_{G+1} = {P_{G+1}} \cup {\Im^l}[1:(\Re - \left| {P_{G+1}} \right|)]$\; 
    $Q_{G+1} \leftarrow$ select, crossover, mutate and evaluate $P_{G+1}$\;
}
\KwRet Pareto front $\Im^{1}$ 
\end{algorithm}

\textbullet \enspace \textit{Constraints checking}: Since the initial population is generated randomly, it is crucial to verify whether each individual $\mathds{I}$ satisfies the constraint $C5$ in problem \eqref{23}, which applies to both the initial population and all subsequently generated offspring. Solutions that do not meet the constraint are deleted and a new value will be generated.

\textbullet \enspace \textit{Performance evaluation}: For each individual, the objective performance matrices $f_m^{obj}$ are evaluated, including average per-device AoI $\bar A$ ($m=1$) and the coverage $cov$ ($m=2$), based on formula \eqref{3} and \eqref{11}. 

\textbullet \enspace \textit{Fast nondominated sorting}: The designed NSGA-II algorithm utilizes a fast nondominated sorting mechanism before the selection process, thereby increasing the likelihood of preserving superior individuals, as illustrated in \textbf{Algorithm 2}. In NSGA-II, if two solutions $\mathds{I}$ and $\mathds{J}$ are compared and $\mathds{I}$ is found to be superior across all objectives than $\mathds{J}$, then $\mathds{I}$ dominates $\mathds{J}$, denoted as $\mathds{I} \prec {\mathds{J}}$. A solution like $\mathds{I}_1$, which is not dominated by any other solution, is classified as a nondominated or Pareto-optimal solution. $n_{\mathds{I}}$ represents the domination count of solutions that dominate $\mathds{I}$, while $H_{\mathds{I}}$ denotes the set of solutions dominated by $\mathds{I}$. The set of all Pareto-optimal solutions forms the Pareto front, also known as the nondominated front. Besides, we define $\Im ^l$ as the set of $l$-th layer of the Pareto front. 
\begin{algorithm}[t]
\small
\renewcommand{\baselinestretch}{0.8}
\caption{Fast nondominated sort}
\LinesNumbered
\KwIn{all individual $\mathds{I} \in {S_{\mathds{I}}}$}
\KwOut{Pareto front $\Im$}
    \For{each individual $\mathds{I} \in {S_{\mathds{I}}}$}{
    $\mathds{I}$ 's domination counter: $n_{\mathds{I}} = 0$, solution set dominated by $\mathds{I}$: $H_{\mathds{I}} = \emptyset$\; 
    \For {other individual $\mathds{J} \in {S_{\mathds{I}}}$}{
    \eIf{$\mathds{I} \prec {\mathds{J}}$}{$H_{\mathds{I}} = H_{\mathds{I}} \cup \mathds{J}$ \quad\quad// add $\mathds{J}$ to set $H_{\mathds{I}}$ \;}{
    \If{$\mathds{J} \prec \mathds{I}$}{$n_{\mathds{I}} = n_{\mathds{I}} + 1$  \quad\quad//increment the counter \;}
    }
    }
    \If{$n_{\mathds{I}} = 0$ } {
        $\mathds{I}^{rank} = 1$, $\Im^1 = \Im^1 \cup \mathds{I}$\;
    }
   } 
   Initialize Pareto front counter $l = 1$\;
   \While{$\Im^l \ne \emptyset$}{
       initialize the next front set $W = \emptyset$\;
       \ForEach{ individual $\mathds{I} \in \Im^l$ }{
       \ForEach{other individual $\mathds{J} \in H_{\mathds{I}}$}{
       $n_{\mathds{J}} = n_{\mathds{J}} - 1$\;
       \If{$n_{\mathds{J}} = 0$}{
       $\mathds{J}^{rank} = l + 1$\; $W = W \cup \mathds{J}$; //\ add $\mathds{J}$ to next front\;
       }}}
       $l = l + 1$\;
       $\Im^l = W$\;
   }
\KwRet Pareto front $\Im$\; 
\end{algorithm}

\textbullet \enspace \textit{Crowded comparison}: For one nondominated set, the crowding-distance $D_i$ of solution ${ \mathds{I}}_i$ is calculated as follows: $D_i=\sum\limits_{m = 1}^2 { ({ f}_m^{obj} [i+1]-{ f}_m^{obj} [i-1])/( f_m^{max}- f_m^{min})}$, where ${f}_m^{obj}[i]$ is the value of the $m$-th objective function for the $i$-th individual, $ f_m^{max}$ and $ f_m^{min}$ are the maximum and minimum values of the $m$-th objective function, respectively. A larger $D_i$ indicates a lower density, suggesting that the front is more evenly distributed, which reflects better diversity retention. Solutions with the smallest and largest function values are assigned an infinite distance value $\infty$. The crowded-comparison operator (${ \prec _n}$) directs the selection process across to ensure a uniformly distributed Pareto-optimal front, defined as:
\begin{equation}   
\begin{array}{l}
{{ \mathds{I}}_1} \ {\prec _n} \ {{ \mathds{I}}_2},{{ \rm {if}  }} \  { \mathds{I}}_1^{rank} < { \mathds{I}}_2^{rank} \ {{\rm {or}}}\\
{{ }}{ \mathds{I}}_1^{rank} < { \mathds{I}}_2^{rank} \ {\& } \ {{{D}}_{{1}}} > {{{D}}_{{2}}},
\end{array}
\end{equation}
where ${\mathds{I}}_i^{rank}$ represents the nondomination rank of ${ \mathds{I}}_i$.

\vspace{0.1cm} 
\begin{table}[!t] 
\setlength{\abovecaptionskip}{1pt}
\renewcommand\arraystretch{1.5}
\centering  %
\footnotesize
\begin{threeparttable}
\vspace{-0.5em}
\caption{ \small Simulation parameters}
\begin{tabular}{c|c}  
\toprule
\textbf{Parameter} & \textbf{Value} \\
\hline  
 Earth's radius & 6371 km \\
 Moon's radius & 1737 km \\
 The unit distance of unit AoI $D_{unit}$ & 3000 km \\
 The given elevation angle constraint ${\theta _c}$  & ${5^ \circ }$ \\
 The $z$-axis amplitude of Halo orbit & 13000 km \\
 Total number of observation points on moon $M$ & 100 \\
 The period of Halo orbit ${T_{halo}}$ & 21284 minutes \\
 Channel BER\footnotemark[3] & $10^{-5}$/$10^{-6}$\\
 Packet size $F_i$ & 1 KByte \\
 Transmission rate $T_{rate}$ & 2 Mbit/s \\
 Observation start time & 1 May 2024 00:00:00 \\
 Observation time & 2 $T_{Halo}$ \\
 Population size\footnotemark[4] $\Re$ & 50/100  \\
 Maximum generation $G_{max}$ & 100 \\
 Crossover rate & 0.8 \\
 Mutation rate & 1 / length(individual) \\
\bottomrule
\end{tabular}
\end{threeparttable}
\end{table}
\footnotetext[3] {We assume that by employing techniques such as modulation, coding, and power control, the channel BER between the lunar satellite and the earth satellite is controlled to $10^{-5}$, while for other links, it is maintained at $10^{-6}$.}
\footnotetext[4] {When $N_{ord} \ge 3$, due to the increased number of variables, we set the population size to 100. In all other cases, the population size is set to 50.}

\textbullet \enspace \textit{Selection, crossover, and mutation}: Individuals with the best fitness values are selected to enter the next generation population using the binary tournament selection method based on the crowed-comparison operator ${ \prec _n}$; The simulated binary crossover method \cite{a53} is employed to recombine the genes of two parent individuals. The crossover probability determines the extent of variation between the offspring's gene values and those of the parents, resulting in two new offspring individuals. Additionally, the polynomial mutation method \cite{a54} is applied, where each individual's genes are mutated according to a specific probability, with the magnitude of mutation governed by a polynomial distribution.

After solving the multi-objective optimization problem \eqref{23} for a given constellation configuration \{$N_{GEO}$, $N_{L1}$, $N_{ord}$, $N_{L2}$\}, we can get the Pareto front of coverage and AoI performance for this constellation. By simulating different constellation configurations, we can compare the performance and aggregate these simulation results, which allows for a comparative analysis to identify the optimal architecture and parameters of different constellations.

\section{Simulation Results}  
\subsection{Parameter Setups}
\begin{table}[t]
\small
\setlength{\belowcaptionskip}{0.2cm} 
\centering \caption{\small The simulation constellation configuration}\label{tab2}
\tabcolsep=0.1cm
		\begin{tabular}{|c|c|c|c|c|c|c|c|c|c|}\hline
       \rule{0pt}{18pt}  %
       $N$ & $N_{GEO}$& $N_{L1}$ & $N_{ord}$ & $N_{L2}$ & $N$& $N_{GEO}$& $N_{L1}$ & $N_{ord}$ & $N_{L2}$  \\  \hline	
   4&1&1&1&1& 6&2&2&1&1\\ 
   5&2&1&1&1 & 6&2&1&1&2\\ 
   5&1&2&1&1& 6&1&2&1&2 \\
   5&1&1&1&2 & 6&2&1&2&1 \\
   5&1&1&2&1&  6&1&2&2&1\\
   6&1&1&2&2 & 6&1&1&3&1 \\ \hline
   7&2&2&1&2& 8&2&2&2&2\\ 
   7&2&2&2&1 & 8&2&2&3&1\\  
   7&2&1&2&2& 8&2&1&3&2 \\
   7&1&2&2&2 & 8&1&2&3&2 \\
   7&2&1&3&1&  8&2&1&4&1\\
   7&1&2&3&1 & 8&1&2&4&1 \\ 
   7&1&1&3&2 & 8&1&1&4&2 \\ 
   7&1&1&4&1 & 8&1&1&5&1 \\ \hline
		\end{tabular}
		\label{tab:Margin_settings}
	\end{table}
In this section, we present and analyze the simulation results of various constellation configurations based on NSGA-II. Typically, aligned with the fourth phase of China’s lunar exploration program, the Chang’e-4 (-45.5$^\circ$S, 177.6$^\circ$E) and Chang’e-6 (-41.64$^\circ$S, -153.98$^\circ$W) missions landed in the southern hemisphere of moon, while Chang’e-7 and Chang’e-8 are expected to establish a basic scientific research station near the lunar South Pole. Consequently, we select an area covering latitudes between $[-40^\circ \rm{S},-90^\circ \rm{S}]$ in the southern hemisphere as the target lunar detection area $S$. That is, the constellation design aims to maximize the coverage of area $S$ of the fourth phase of the Chinese exploration missions while improving the information timeliness of the future full moon exploration. The values of simulation parameters are shown in Table II.

Here, we reintroduce the constellation configuration structure \{$N_{GEO}$, $N_{L1}$, $N_{ord}$, $N_{L2}$\}. Based on our previous constellation analysis, the symmetry of the Halo orbits allows for near-complete coverage of LNS (LFS) with one EML1 (EML2) northern family and one southern family Halo orbit satellite. Therefore, we restrict $N_{L1}$ and $N_{L2}$ to a maximum of 2 and set $N_{GEO}$, $N_{L1}$, $N_{ord}$, and $N_{L2}$ to at least 1. When only one satellite is deployed in the EML1 or EML2 Halo orbit, we select the southern Halo orbit. For two satellites, one is placed in the southern Halo orbit and the other in the northern Halo orbit. The amplitude of the Halo orbit is set to 13000 km, as our previous constellation design studies \cite{a43, a44} have demonstrated that satellite in this amplitude value provides excellent coverage performance while avoiding the lunar masking phenomenon. Once the network structure is determined, the EML1/L2 Halo orbit parameters are determined. 

For the EGSs, we select four Chinese deep space tracking and control EGSs located in Xinjiang (38.43$^\circ$N, 76.71$^\circ$ E), Beijing (40.56$^\circ$N, 117$^\circ$ E), Kunming (25.03$^\circ$N, 102.8$^\circ$E), and Heilongjiang (46.50$^\circ$N, 130.78$^\circ$E). During each observation period, we will sample one group orbital and EGSs positions per hour for a total of 711 orbital points, and then calculate $\bar A$ and $cov$. If the source node cannot communicate with EGSs, we set the AoI to a very large value, 2000. Through geometric analysis, when the constellation includes only one GEO satellite, there may be instances where both EGSs and GEO are away from the moon, earth's occlusion would prevent the EGSs from establishing communication links with the moon. However, when there are two GEO satellites, as long as their longitude difference is more than $20^\circ$, at least one GEO satellite will be visible to the moon and EGSs, ensuring continuous communication between EGSs and the moon, so we restrict $N_{GEO}$ to a maximum of 2. To simplify the optimization problem, we set $\xi_2 =\xi_1 + 60^\circ$ if there are two GEO satellites, then GEO satellites require only one variable $\xi_1$.

As for the ordinary lunar orbit parameters $\Phi_k=\{a_k, e_k, i_k, \Omega_k, \omega_k, \nu_k\}$, since we are using circular orbits, both $e_k$ and $\omega_k$ are 0. Additionally, the entire constellation must satisfy the period constraint $C8$ in equation \eqref{23}. To meet this, we first calculate the possible orbital periods $T_{ord}$ for circular orbits based on the Halo orbit period and equation \eqref{9}. From this, we derive the corresponding semi-major axis $a$ using equation \eqref{10}. The final values of $a$ that satisfy constraint $C5$ in equation \eqref{23} are: 2650, 3210, 3525, 5596, 8882, and 14100 km. If there are one circular orbit, the optimization variables are $\{a, i_1, \Omega_1, \nu_1, \xi\}$. For every one unit increase in $N_{ord}$, the variable increases $\{i, \Omega, \nu \}$, i.e., the variables are $\{a, i_1, \Omega_1, \nu_1,i_2, \Omega_2, \nu_2, \xi\}$ for $N_{ord}=2$. In this paper, we will simulate the constellation structure configurations of different $N$ as shown in Table III.

\begin{figure*} [t]
        \center
        \scriptsize
        \begin{tabular}{ccc}
                \includegraphics[width=5.8cm]{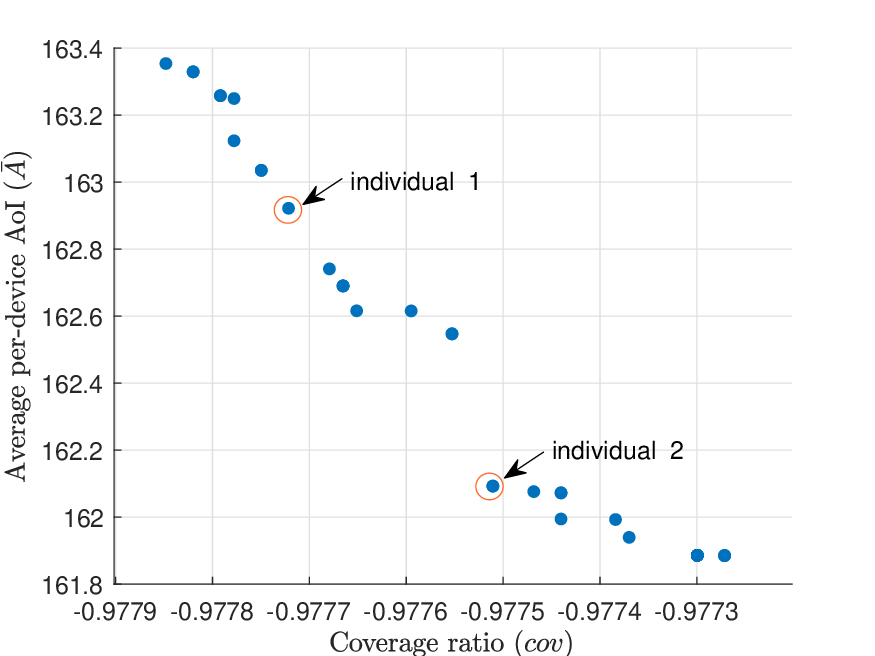} & \includegraphics[width=5.8cm]{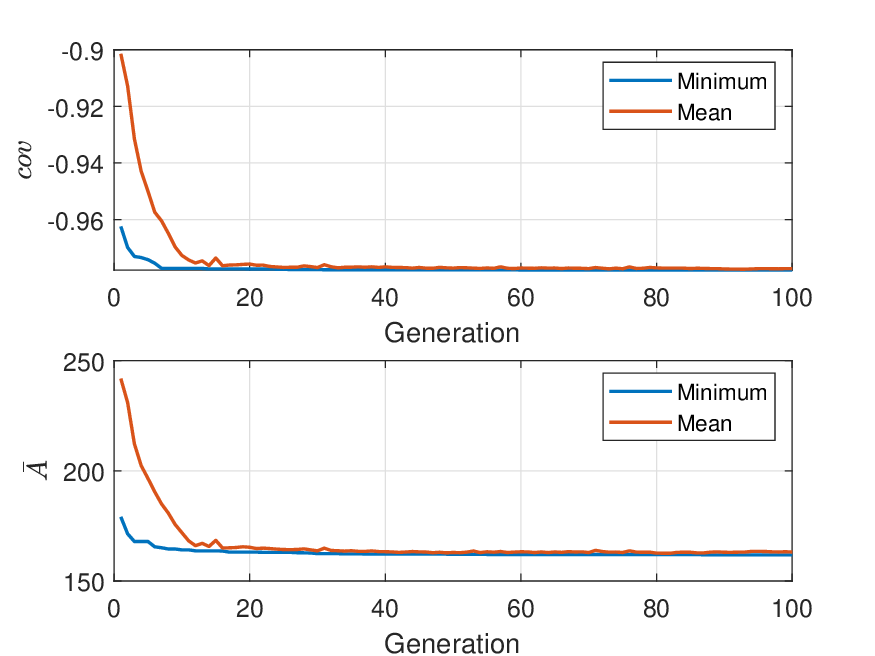} & \includegraphics[width=5.8cm]{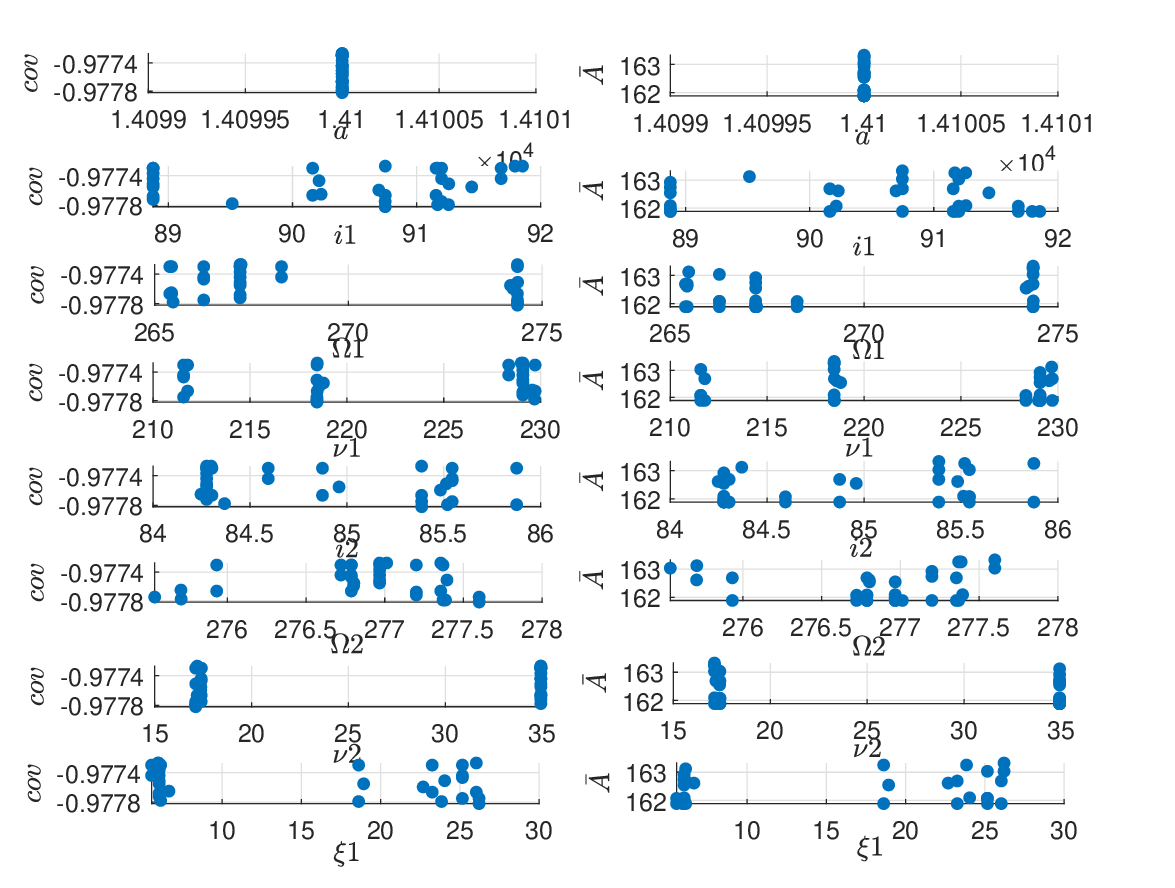}\\
                (a): Pareto front &  (b): Objective evolution over generations &(c):Variables vs. objectives
        \end{tabular}
\captionsetup{font=small,labelsep=period, singlelinecheck=false}
        \captionsetup{labelformat=modified, justification=raggedright}
        \caption{\small Pareto front and convergence of NSGA-II for constellation configuration \{2,1,2,1\}.}
        \label{fig7}
        \vspace{-0.5em}
\end{figure*}
\begin{figure*} [!ht]
        \center
        \scriptsize
        \begin{tabular}{ccc}
                \includegraphics[width=5.8cm]{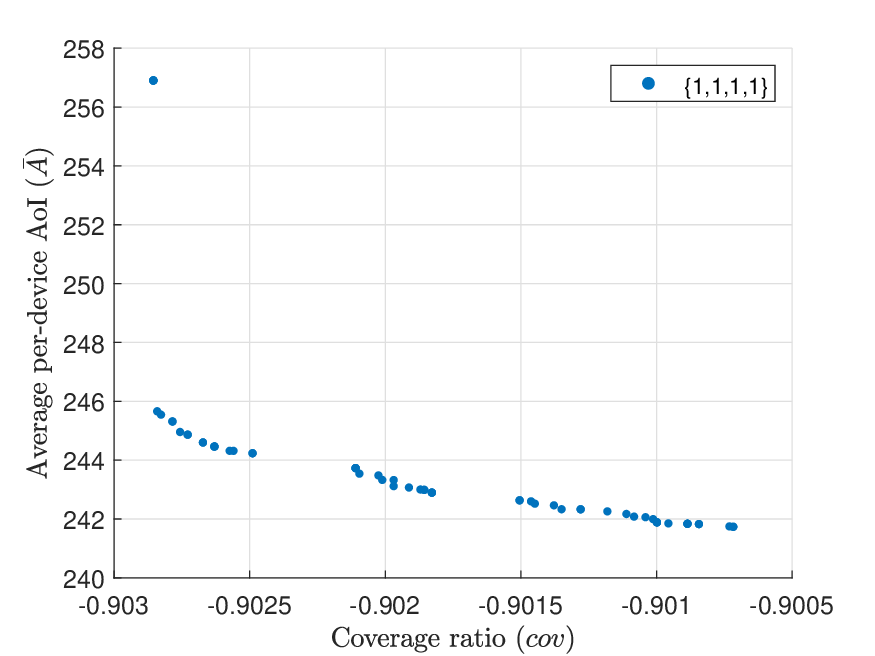} & \includegraphics[width=5.8cm]{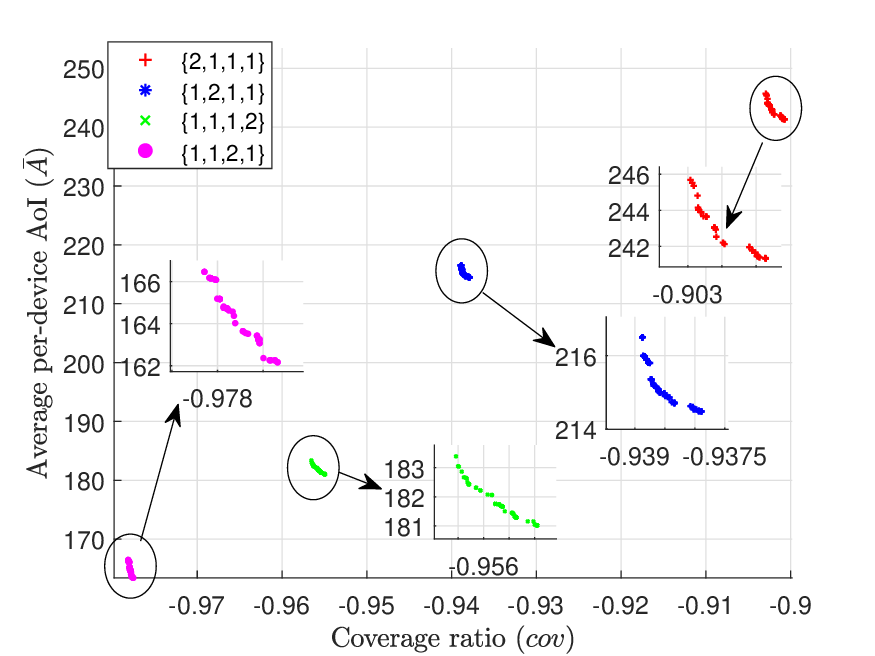} & \includegraphics[width=5.8cm]{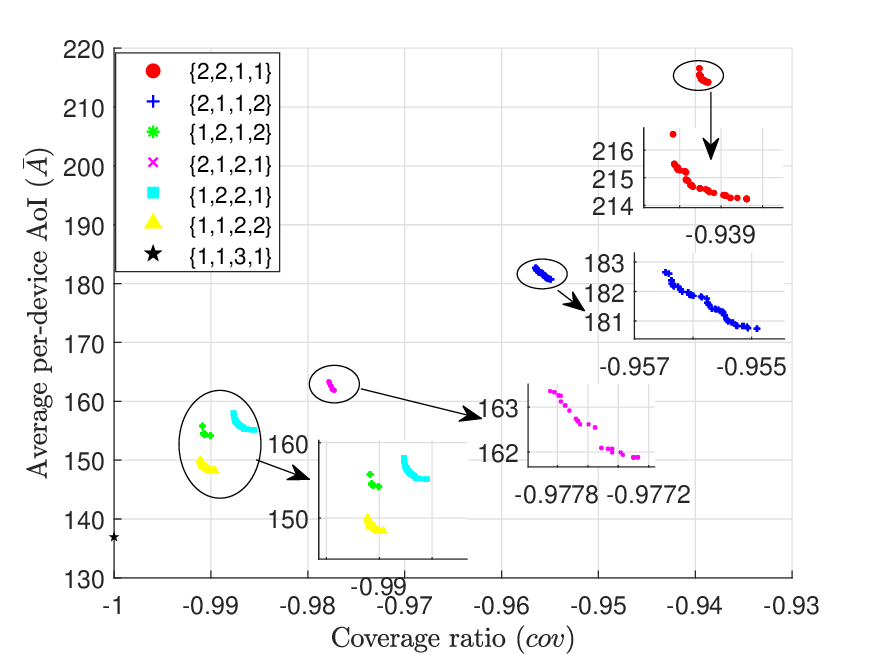}\\
                 (a): $N=4$ &   (b): $N=5$ &(c): $N=6$ \\
               \includegraphics[width=5.8cm]{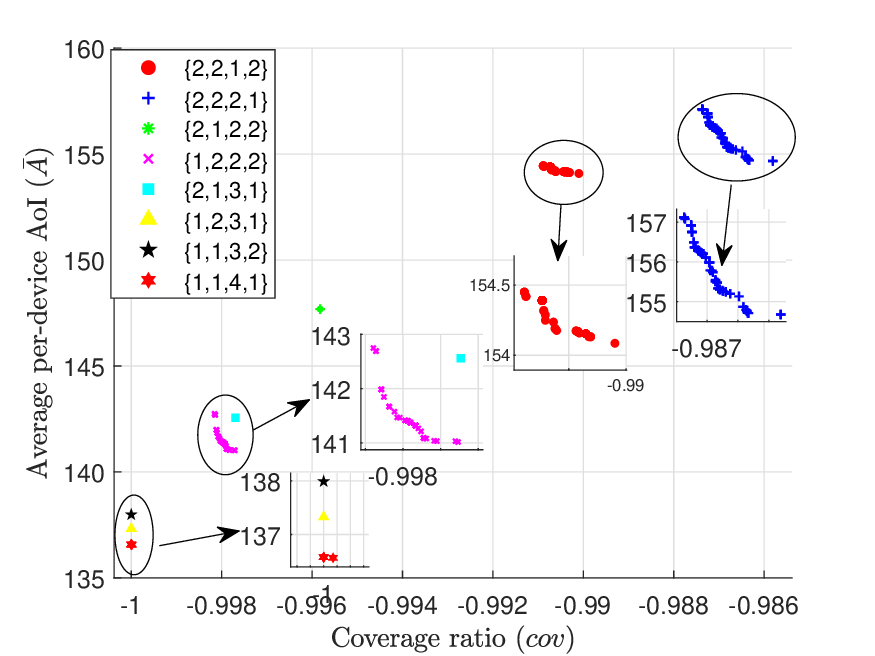} & \includegraphics[width=5.8cm]{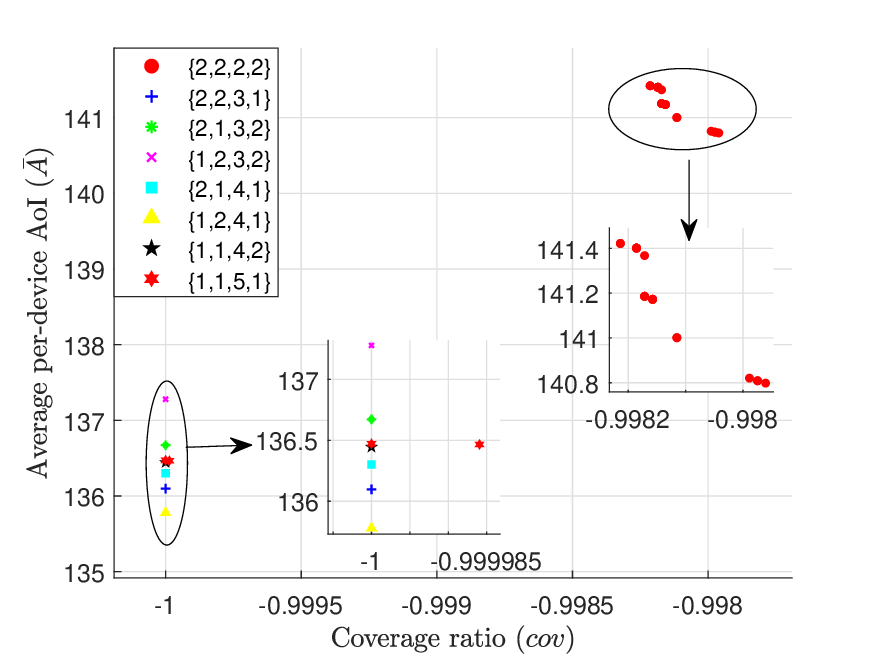} & \includegraphics[width=5.8cm]{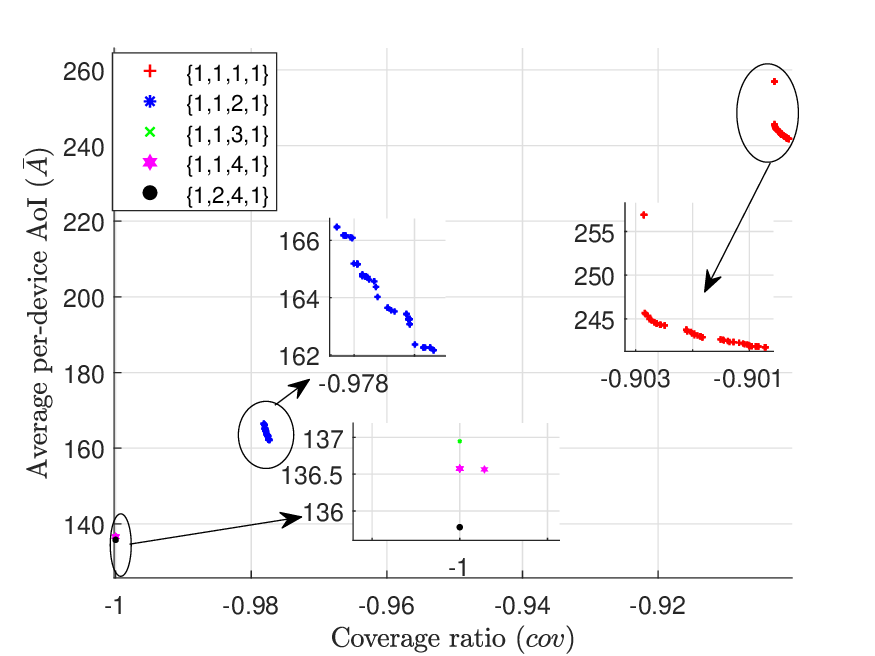} \\
                (d): $N=7$ &    (e): $N=8 $&    (f): Optimal Pareto front of different $N$   \\
        \end{tabular} \captionsetup{font=small,labelsep=period, singlelinecheck=false}
        \captionsetup{labelformat=modified, justification=raggedright}
        \caption{\small Pareto front comparison of different network structures.}
        \label{fig8}
        \vspace{-0.5em}
\end{figure*}
\subsection{Numerical Results}
We first simulated one possible configuration \{2,1,2,1\}. Fig. \ref{fig7} shows the Pareto front and the best solutions obtained by the NSGA-II algorithm, the convergence of the NSGA-II algorithm, as well as the relationships between variables and objectives for the Pareto front solutions. Fig. \ref{fig7}(a) is the Pareto front, we can select the appropriate individual based on the threshold requirements for the objectives $cov$ and $\bar A$. For example, if we want $cov$ to be no less than 97.75\%, we can select the individual 2 that meets the coverage requirement and has the smallest $\bar A$, and if we want $cov$ to be no less than 97.77\%, we can select individual 1. Fig. \ref{fig7}(b) shows the evolution of objectives over generations, NSGA-II gradually approaches the optimal value, and coverage and AoI both achieve convergence in about 20 generations. Therefore, the proposed algorithm has good convergence. Fig. \ref{fig7}(c) shows the relationship between variables and objective function. According to the optimization results, the inclinations $i1$ and $i2$ of the two circular lunar orbit satellites are about 90°, and the right ascension of ascending Node $\Omega1$ and $\Omega2$ are about 270°, that is, the circular orbits are at the junction of LNS and LFS, and the phase difference between the two satellites is $\nu 1 - \nu 2 \approx {180^ \circ }$. Because one EML1 and one EML2 Halo orbit satellite can almost cover $S$, the uncovered area is the junction of LNS and LFS, the optimized circular orbit satellite can cover most of the uncovered region, which is completely consistent with the concept of our constellation design.

Fig. \ref{fig8} compares the Pareto fronts of various network configurations under different total satellite numbers, $N$. The closer the Pareto frontier is to the lower left corner, the better AoI and coverage performance will be. Fig. \ref{fig8}(a) illustrates the Pareto front of $N=4$ (configuration \{1,1,1,1\}). Approximately 90\% coverage of the target area $S$ is achieved, with an average AoI $\bar{A}$ ranging between [241, 257]. This value is significantly larger than 128 (computed as 384400/3000), indicating that not all lunar observation points can communicate with the EGSs in real-time. Fig. \ref{fig8}(b) presents the Pareto fronts for four different constellation configurations when $N=5$: \{2,1,1,1\}, \{1,2,1,1\}, \{1,1,1,2\}, and \{1,1,2,1\}. Among these, the structure \{1,1,2,1\} demonstrates the best performance in terms of both AoI and coverage, achieving an average $\bar{A}$ of approximately 164 and $cov$ of 97.8\%. The next best configuration is \{1,1,1,2\}, which attains a $\bar{A}$ of 182 with a $cov$ of 95.6\%. The configuration \{1,2,1,1\} follows, with $\bar{A}$ of approximately 215 and $cov$ of 93.85\%, while the \{2,1,1,1\} structure provides $\bar{A}$ of 244 and $cov$ of 90.2\%. When the total number of satellites, $N$, is constrained, placing more satellites on the lunar side leads to greater coverage of the lunar surface, thereby reducing $\bar{A}$ and increasing $cov$. Fig. \ref{fig8}(c) presents the comparison for $N=6$. The configuration \{1,1,3,1\} shows the best AoI and coverage performance, achieving 100\% coverage and an AoI of approximately 137. The next best configuration is \{1,1,2,2\}, which achieves AoI of 148 and coverage of 99\%. This is followed by configurations \{1,2,1,2\}, \{1,2,2,1\}, \{2,1,2,1\}, \{2,1,1,2\} and \{2,2,1,1\}, with the latter showing the weakest performance. For smaller $N$, a single GEO satellite demonstrates better performance and more satellites should be positioned on the lunar side to reduce $\bar A$ and increase $cov$. Therefore, configurations such as \{2,2,1,1\}, \{2,1,1,2\}, and \{2,1,2,1\} perform slightly worse due to less optimal satellite placement. By comparing the configurations \{1,1,2,2\}, \{1,2,1,2\}, and \{1,2,2,1\}, it becomes clear that the structure with more satellites closer to the LFS has better performance. Fig. \ref{fig8}(d) compares the results for $N=7$. Configurations \{1,1,4,1\}, \{1,1,3,2\}, and \{1,2,3,1\} all achieve full coverage ($cov = 100\%$), with \{1,1,4,1\} showing the best overall performance. Fig. \ref{fig8}(e) presents the results for $N=8$, where all configurations, except \{2,2,2,2\}, provide full coverage of $S$. The constellation \{1,2,4,1\} exhibits the best performance among them. Lastly, Fig. \ref{fig8}(f) compares the optimal constellations for various values of $N$. It is evident that for $N=6$, $N=7$, and $N=8$, full coverage of $S$ is achieved, with only slight differences in AoI. Overall, as $N$ increases, performance improves. Among the configurations, \{1,1,3,1\} demonstrates the best performance when $N$ is limited. 
\begin{figure}[t]
\centering
\includegraphics[width=0.47\textwidth]{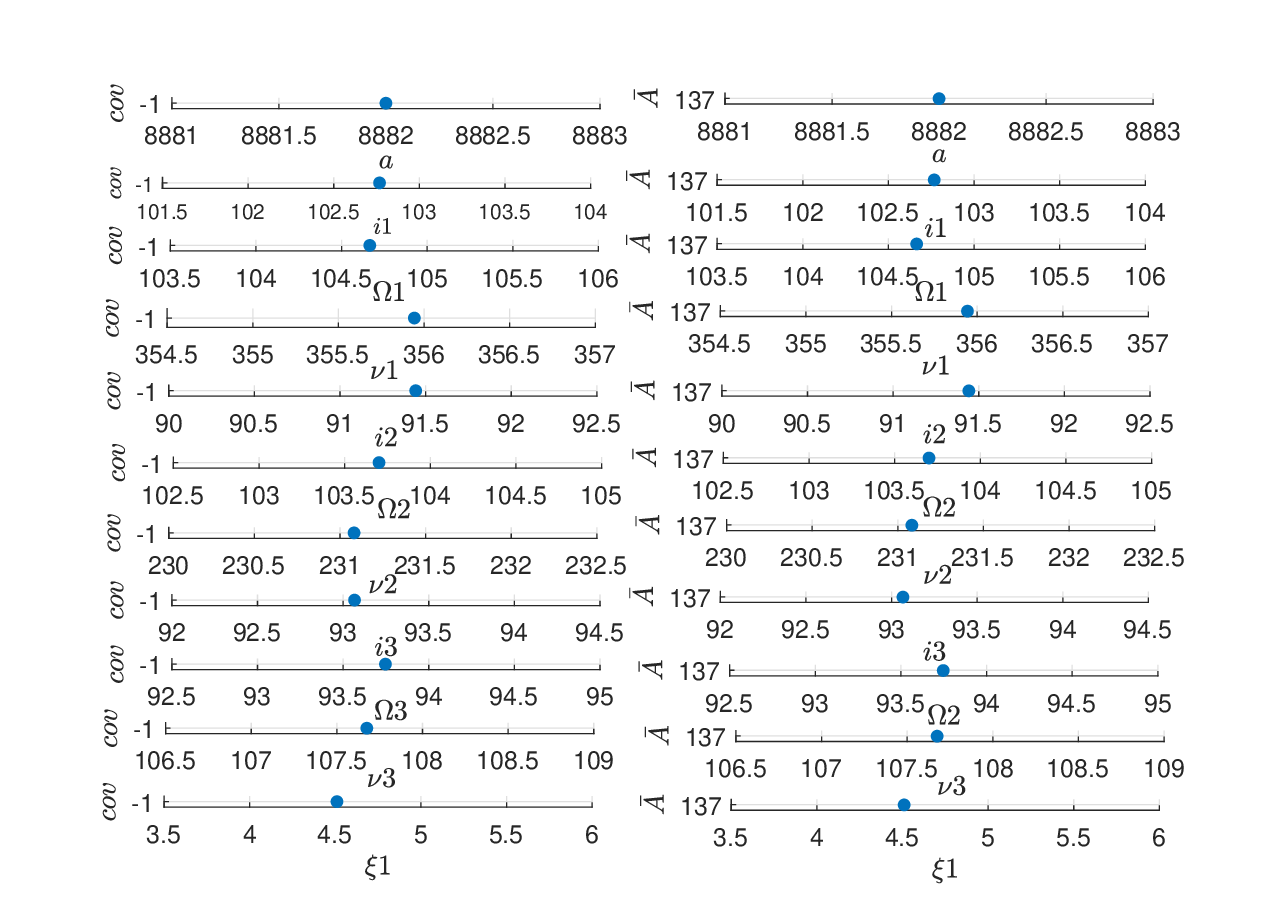}
\captionsetup{font=small,labelsep=period, singlelinecheck=false}
\captionsetup{labelformat=modified, justification=raggedright}
\caption{\small  Variables vs Objectives in Network \{1, 1, 3, 1\}.} \label{fig9}
\end{figure}

Fig. \ref{fig9} illustrates the relationship between variables and objective functions under the network configuration \{1, 1, 3, 1\}. It can be observed that for the three satellites, the orbital inclinations are approximately 90° for two satellites and about 104° for one satellite. The respective phase angles $\nu 1$, $\nu 2$, $\nu 3$ are approximately 356°, 231°, and 108°, with a semi-major axis $a$ of 8882 km. The three satellites are situated near the intersections of LNS and LFS, with their phase differences being almost uniformly distributed. This configuration further validates the constellation design approach of using circular orbit satellites to compensate for coverage gaps in EML1/L2 Halo orbit satellites.

\begin{figure}[t]
\centering
\includegraphics[width=0.47\textwidth]{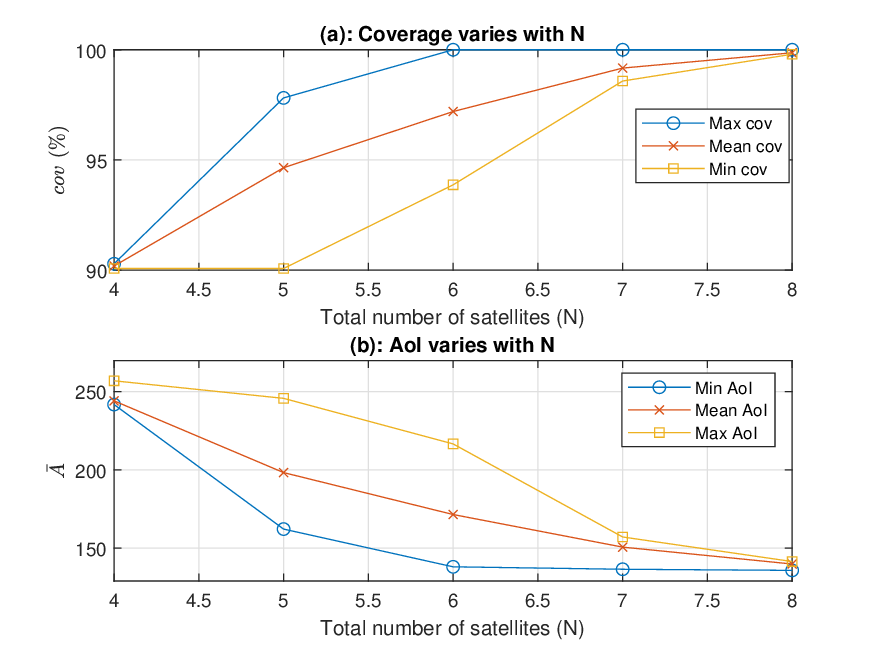}
\captionsetup{font=small,labelsep=period, singlelinecheck=false}
\captionsetup{labelformat=modified, justification=raggedright}
\caption{\small  Coverage and AoI change with total number $N$.} \label{fig10}
\end{figure}
We extract the final Pareto fronts of different configurations for various total number of satellites ($N$) and analyzed the maximum, minimum, and average values corresponding to the Pareto fronts of different configurations. Fig. \ref{fig10} illustrates the changes in coverage and AoI performance as $N$ increases. From Fig. \ref{fig10}(a), it can be observed that the coverage performance of the target area $S$ improves as $N$ increases, achieving full coverage when $N=6$. As $N$ increases further, both the minimum and average coverage rates approach 100\%. Fig. \ref{fig10}(b) shows the variation of AoI with $N$. As $N$ increases, the AoI $\bar A$ gradually decreases. When $N=6$, the minimum AoI does not decrease significantly as $N$ increases to 7 and 8, but the average and maximum values decrease considerably. This indicates that with an increase in $N$, the AoI performance gradually improves, and the overall performance shows substantial enhancement.

\begin{figure}[t]
\centering
\includegraphics[width=0.47\textwidth]{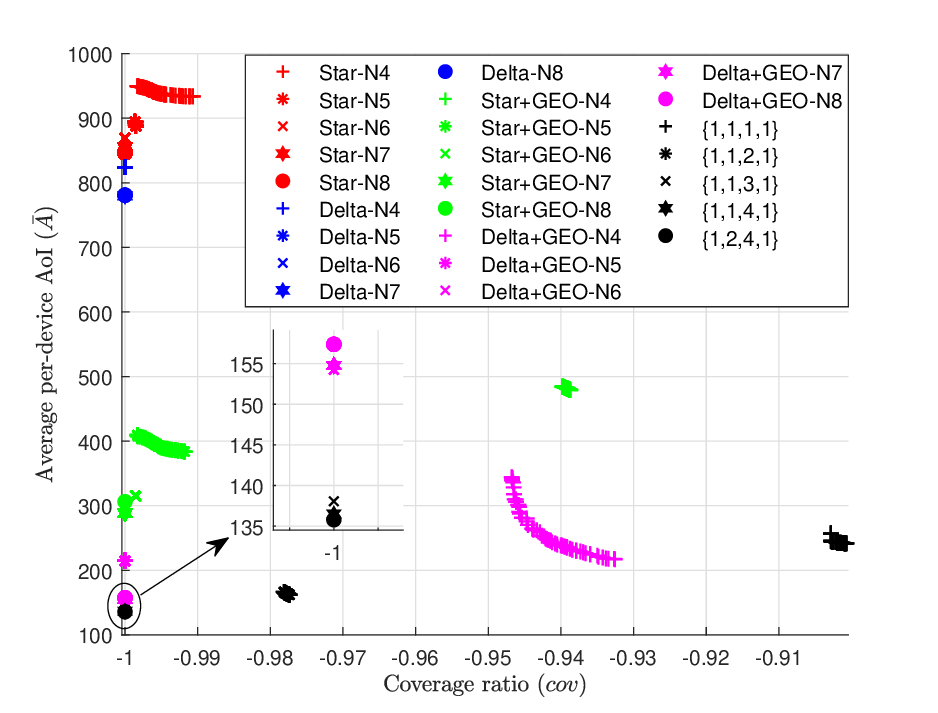}
\captionsetup{font=small,labelsep=period, singlelinecheck=false}
\captionsetup{labelformat=modified, justification=raggedright}
\caption{\small Pareto front comparison of proposed and different Walker constellations.} \label{fig11}
\end{figure}

Fig. \ref{fig11} compared the Walker Star constellation (with an inclination of $i=90°$) and the Delta constellation (with $0° < i < 90°$) for different total satellite numbers, $N$. The $N$ satellites are evenly distributed in $N$ planes, and the Right Ascension of Ascending Node $\Omega$ and Argument of the Perigee $w$ are evenly distributed. For detailed configurations, refer to Section II of reference \cite{a27}. First, we simulated the performance of the Star constellation without GEO satellite, with all satellites located on the lunar side, as shown in red lines. Although the performances improve as $N$ increases, the AoI remains high, reaching over 800. The blue lines represent the Delta constellation without GEO satellite, which performs better than the star constellation due to its adjustable inclination. It is evident that the traditional Star and Delta constellations focus solely on lunar-side coverage without accounting for the earth's rotation and the moon's orbital motion. They achieve good coverage on the lunar side, but they do not offer high timeliness for Earth-Moon communications. Additionally, we simulated the scenarios where a GEO satellite is included. One satellite from $N$ is relocated to the earth side to address both the earth's rotation and the moon's orbit. The simulation results show that the Star+GEO and Delta+GEO constellations significantly outperform traditional constellations in terms of both AoI and coverage performance. In addition, the performance of our proposed constellations in black lines is significantly better than the traditional Walker Star and Delta constellations. For example, when \(N=6\), under similar $cov$ performance, the $\bar A$ of the proposed optimal constellation $\{1, 1, 3, 1\}$ improves by 11.04\%, 56.55\%, 82.45\%, and 84.25\% compared to Delta+GEO-N6, Star+GEO-N6, Delta-N6, and Star-N6, respectively. Therefore, our proposed combined Earth-Moon relay satellite constellation will greatly enhance the freshness of system data.
\section{Conclusion}  
In this paper, we propose a heterogeneous Earth-Moon relay network comprising GEO, circular lunar orbit, and EML1/L2 Halo orbits satellites to enhance the information freshness and improve the coverage ratio of the area near the lunar south pole using the minimal number of satellites. This problem is modeled as a multi-objective optimization problem, which is solved using the NSGA-II algorithm. By simulating constellations with different configurations, we obtain the optimal constellation design under a limited number of satellites. The results show significant performance improvements compared to the traditional Walker Star and Delta constellations, validating the effectiveness of our constellation design, which will contribute significantly to advancements in space communication technologies and lunar exploration efforts. In future work, we will explore the design of a joint relay constellation and user routing strategy to further enhance data timeliness within the Earth-Moon communication system, particularly for a densely populated, key area on the lunar surface.

\end{document}